\documentclass[journal]{IEEEtran}

\usepackage{amssymb}
\usepackage{amsthm}
\usepackage{amsmath} %
\usepackage{mathtools}
\usepackage{dutchcal} 
\usepackage{bm}
\usepackage{graphicx}
\usepackage{epstopdf}
\usepackage{subfigure}
\usepackage{cite,color}

\definecolor{red}{rgb}{1,0,0}  

\newcommand{\expect}[1]{\mathbb{E}\left\{#1\right\}}   
\newcommand{\expectl}[1]{\mathbb{E}\{#1\}}   

\newcommand{\gcal}{\mathcal{g}}
\newcommand{\Hcal}{\bm{\mathcal{H}}}
\newcommand{\hcal}{\bm{\mathcal{h}}}
\newcommand{\Ical}{I}

\newcommand{\Scal}{S}

\newcommand{\Hbf}{{\bf H}}
\newcommand{\hbf}{{\bf h}}

\newcommand{\Rbf}{{\bf R}}

\newcommand{\sbf}{{\bf s}}

\newcommand{\Vbf}{{\bf V}}
\newcommand{\vbf}{{\bf v}}
\newcommand{\Wbf}{{\bf W}}
\newcommand{\wbf}{{\bf w}}

\newcommand{\Nt}{N_{\rm t}}

\newcommand{\Nu}{N_{\rm u}}

\newcommand{\Nsc}{N_{\rm sc}}
\newcommand{\Nfft}{N_{\rm fft}}
\newcommand{\Ncp}{N_{\rm cp}}
\newcommand{\Hrm}{\dag }
\newcommand{\Trm}{{\rm T}}

\newcommand{\jrm}{{\rm j}}
\newcommand{\Th}{\tau_{\rm tr}}

\newtheorem{Cor}{Corollary}

\newtheorem{Prop}{Proposition}
\newtheorem{Lem}{Lemma}

\begin{document}

\title{
Channel Shortening by Large Multiantenna Precoding in OFDM 
}

\author{
\IEEEauthorblockN{Renaud-Alexandre Pitaval}
\thanks{
The author is with Huawei Technologies Sweden AB (email: renaud.alexandre.pitaval@huawei.com).

Part of this work was presented at the 2018 IEEE Global Conference on Communications~\cite{PitavalGlobecom18}.
}
}

\maketitle

\begin{abstract}

A channel delay spread  larger than the cyclic prefix (CP)  creates inter-carrier/symbol interference (ISI/ICI) in orthogonal frequency-division multiplexing (OFDM). Recent interests in low-latency applications have motivated the usage of shorter OFDM symbols where one can either downscale the CP at the cost of interference, or maintain it but with  larger overhead. 
Alternatively, this paper studies channel shortening methods exploiting the properties of large multi-antenna precoding  
in order to steer the transmitted signal energy toward channel paths inside an insufficient CP. 
It is shown that ISI/ICI can asymptotically be   canceled by conventional subcarrier-based precoding with an infinite number of antennas. 
This is achieved by introducing time-delay selectivity inside frequency-selective precoders in order to remove undesired delayed signals, providing a trade-off  between interference mitigation and multi-path combining gains, and leading to subsequent gains in high SNR.    
This frequency-domain precoding method, coined time-frequency (TF) precoding, is compared to time-reversal (TR) filtering  
whose  asymptotic rate is optimal but introduces post-modulation processing with channel-dependent signal distortion.  
In addition to maintain the legacy precoded multi-antenna  OFDM structure, finite-size analysis shows that TF-precoding converges faster to its asymptotic rate than TR-filtering, so that TF-precoding can outperform TR-filtering in the  high-SNR regime with not-so-many antennas.   
\end{abstract}


\section{Introduction}
\label{s:introduction}

Orthogonal frequency division multiplexing (OFDM) is to date the dominant modulating waveform of wireless communications systems. When combined with a long enough cyclic prefix (CP), OFDM transforms a multi-tap channel in time into multiple orthogonal single-tap subchannels in frequency. One of the features that made OFDM undefeatable  so far by any other waveform technology  is its simple combination with multiple-antenna systems.  
Multiple antennas at transceiver nodes can notably boost the signal-to-noise-ratio (SNR) of transmission links via precoding gains. For broadband systems, frequency-selective multi-antenna precoding in OFDM can simply be designed and applied on each subcarrier;   
a simple combination which has been of paramount importance in the success of this system.  As a result, a large majority of multi-antenna precoding designs and analyses in the literature is based on a narrowband channel assumption with analogy to a single subcarrier transmission where the channel is implicitly the Fourier transform of a channel impulse response (CIR) evaluated at a targeted subcarrier frequency. 

It is foreseen that for the next-generation wireless communication networks to enable new usages, they should not only provide an expected increase of data-rate but also achieve much lower latencies and higher reliabilities.  
To satisfy more stringent user latency requirements, the transmission time duration should be shortened, which can be achieved by  consuming more frequency resources, i.e., using a larger subcarrier spacing (SCS) in OFDM. Downscaling proportionally  the CP might then increase inter-symbol and  inter-carrier interference (ISI/ICI). Alternatively, the CP time duration can be maintained but at the cost of an increased overhead.  
For example and toward this direction, the recent 3GPP 5G NR standard~\cite{3GPPTS38.211} introduces several new OFDM numerologies with larger SCSs than those in the 4G LTE~\cite{3GPPTS36.211}: e.g., 
 30 and 60 kHz are  supported for  sub-6GHz spectrums in addition to the legacy LTE 15 kHz SCS.  
The \emph{normal CP} duration is accordingly  downscaled at a constant 7\% CP overhead, but these new CP lengths can be shorter than the delay spreads of some channel models as previously used in 3GPP evaluations. Therefore in order to cope with this, 5G NR introduced in addition an \emph{extended CP} for 60 kHz SCS~\cite{3GPPTS38.211}, which however corresponds instead to a high  25\% CP overhead. 

A key ingredient to boost the data rate of new generation networks is to equip base stations (BSs) with large antenna arrays. Indeed, precoding gain scales up with the number of antennas unlike several possible impairments, such as multi-user interference and channel estimation error, which in turn relatively vanish~\cite{MarzettaConf,Marzetta}.  
Using the same observation, it was shown in~\cite{pitarokoilis2012optimality} that single-carrier transmission becomes optimal when combined with time-reversal (TR) filtering and an infinite number of antennas. 
This method could as well be applied  to OFDM but then would not exploit one of the main features for which OFDM has been designed and praised for: namely, equalization and precoding by multiplications in the frequency-domain rather than by time-convolution. So far, it has been rather unclear  in the literature if large-antenna systems could remove ICI/ISI with a typical multi-antenna OFDM system implementation, i.e. where  precoders are applied individually on each subcarrier. 
Still, one can  remark that there have been some intuitive clues supporting this. 
First, a reasoning sometimes encountered in engineering circle is: ``because 
the effective precoded channels of all  subcarriers are asymptotically  tending to the same constant as shown in~\cite{Marzetta}, the overall frequency-selectivity is vanishing and thus by reciprocity one can deduce that the channel time-dispersion caused by multi-path propagation is disappearing.''  However, this is only partially-correct as the model in~\cite{Marzetta} already implicitly assumes a longer CP than the delay spread, and one thus cannot properly draw a conclusion  on a  potential CP reduction from this observation. In passing, we will formalize and clarify in Lemma 1 of this paper the exact asymptotic time-dispersion of such effective channel.    
Secondly, it is also intuitively understood that large multi-antenna precoding can produce  asymptotically narrow beams, and when  applying directional beamforming to, e.g., a dominant line-of-sight channel, the transmission should somewhat reduce to a single-tap channel.  
 
Nonetheless, 
it was observed in~\cite{NsengiyumvaMT} by simulations of conventional large-antenna precoding that interference increases if shortening the CP while at the same time increasing the number of antennas. Still, it was suggested that the CP  could slightly be  reduced for large-antenna systems for increasing the rate. 
A significant clarification was provided in~\cite{aminjavaheri2017ofdm} by analytically showing that ISI and ICI do not vanish for an OFDM transmission without CP when using conventional precoders with an infinite number of antennas, and thus, contrary to when using a sufficient CP~\cite{MarzettaConf},  the SINR saturates as the number of antennas grows. It was also shown that potential multi-user interference does not play a role in this SINR saturation  resulting from an insufficient CP, and the same conclusion would follow for channel estimation error as  in~\cite{Marzetta}.  The SINR saturation problem in~\cite{aminjavaheri2017ofdm} was then circumvented  by using instead TR-filtering with OFDM. 

TR-filtering has recently received  academic interests and  we refer to~\cite{aminjavaheri2017ofdm} for a good literature review on this. TR-filtering is  a time-domain equalization (TEQ)   
whose asymptotic optimality relies on the properties of large multi-antenna arrays. 
For a single-antenna system, such TEQ method may be far from optimal and even degrade the performance by increasing further the time-dispersion of the channel.   
One may note that many other TEQ methods optimizing different criteria have been investigated in the context of single or few antenna systems, see e.g.,~\cite{HuaPhD,TEQ_TSP05} and references therein.  
While CP-OFDM has been in part devised to circumvent TEQ by frequency-domain equalization at the receiver, TEQ has still been considered for OFDM in the case of an insufficient CP~\cite{TEQ_TSP05}. In this case, it is enough to shorten the effective channel to be inside the CP and TEQs have then often been referred to as  \emph{channel shortening} methods.   
At the transmitter-side, TEQ  was traditionally not considered because of the cost of acquiring channel state information (CSI), an issue that has become less relevant since the emergence of precoded multi-antenna systems which heavily rely on CSI at the transmitter;  but also because of other  concerns such that, e.g., undesirable signal distortion and  frequency-division multiplexing.  
 
In this paper, we are instead interested in the possibility of a channel shortening technique realized only by frequency-domain per-subcarrier precoding 
as conventionally done in legacy multi-antenna OFDM systems,  and we will show that large antenna arrays at the BS renders this possible.  
For an OFDM system with an insufficient CP, the strategy is to design frequency-domain precoders that steer the transmitted signal energy principally towards the directions of the channel taps inside the CP.
More specifically and  unlike what might be extrapolated from~\cite{aminjavaheri2017ofdm}, 
an appropriately-modified maximum-ratio-transmission (MRT) precoding  is shown to average out the ISI and ICI in OFDM with an insufficient CP as the number of antennas goes to infinity. 
This is achieved by exploiting the spatial-selectivity offered by large antenna arrays to construct multi-antenna precoders which are time-delay selective in addition to their conventional frequency-selectivity, therefore coined as time-frequency (TF) precoding.  The precoders are then based on the Fourier response of a truncated CIR  of about the CP length, and thus a combination of only certain channel paths that fall within the supported delay range. 

Throughout the paper,  we focus on a single-user downlink data transmission  with  perfect CSI at the BS.
Nevertheless, the analysis and results are also directly applicable to the uplink, and  we show through simulations that the TF precoding design can directly be extended to other types of precoders such as zero-forcing (ZF) in a multi-user scenario. 
We remark  also that in a practical setting having a sufficient CP during the uplink training phase may be necessary to obtain an accurate CSI at the BS, which may then be  used for the downlink data transmission by exploiting the uplink-downlink channel reciprocity.

TF-precoding  is studied via asymptotic rate analysis as the number of BS antennas grows large. It shows that it may bring subsequent gains in the high SNR-regime compared to a conventional frequency-domain precoding (F-precoding) without time-delay selectivity. The selection of the truncation threshold in TF-precoding that asymptotically maximizes the rate  corresponds to an optimization trade-off between interference mitigation and multi-path combining gain. 
TF-precoding is also compared to TR-filtering which instead relies on channel-dependent time-domain processing, distorting the signal in a less controlled and thus potentially undesirable manner, but  is asymptotically optimal and therefore provides an interesting  upper bound.    
  
Asymptotically-tight rate approximations for a finite number of antennas are also derived for both TF-precoding and TR-filtering. These   reveal that TF-precoding converges faster to its asymptotic performance than TR-filtering with and without CP. As a result, TF-precoding  
might even   outperform TR-filtering with not-so-many antennas in the high-SNR regime. The analysis is further confirmed by numerical symbol error rate (SER) and throughput evaluation with a finite array system of 64 and 200 antennas.  

In Section II, the OFDM system model with multi-antenna precoding is presented%
\footnote{\emph{Notation:} Matrices and vectors are
denoted by boldface uppercase and lowercase variables, respectively. Frequency-domain channel variables are in addition distinguished by a calligraphic font as, e.g., $\hcal$, $\gcal$, and $\Hcal$.  
The superscripts $(\cdot) ^*$, $(\cdot)^\Trm$,  and $(\cdot)^\Hrm$ denote complex conjugate, transpose, and conjugate transpose, respectively.
The linear convolution between discrete functions $f[n]$ and $g[n]$ is denoted by $(f*g)[n]$, $\delta_{i,j}$ is the Kronecker delta function, and $\expect{\cdot}$ is the expected value of a random variable.  
}.  
In Section III, the alternative system model with TR-filtering is presented. Section IV provides an asymptotic rate analysis with precoder optimization and numerical comparisons. Section V provides a finite-size rate analysis, while Section VI provides corresponding link-level simulations. Section VII discusses briefly some extensions of this work, and Section VIII concludes the paper.    

\section{Precoded Multi-Antenna OFDM}
\label{s:SysModel} 
We consider a BS transceiver with a large number of $\Nt$ antennas communicating to a user equipment with a single antenna,  and concentrate on a downlink data transmission with perfect CSI at the BS.  

\subsection{System Model}
An OFDM modulation is defined by a subcarrier spacing $\Delta_f$, an inverse fast Fourier transform (IFFT) of size $N_{\rm fft}$, and a CP length $\Ncp$.  An OFDM symbol (without CP) has then a time duration of $T_{\rm s}= 1/\Delta_f$ with sampling period $T_{\rm sp}=T_{\rm s}/ N_{\rm fft}$. A total of $\Nsc \leq N_{\rm fft}$ consecutive subcarriers are allocated with  independently and identically
distributed (i.i.d.) data symbols from a zero-mean complex normal distribution  with average power $P$. 

\subsubsection{Transmission}
The data symbol $x_{b,l}$ for the $l$th subcarrier of the $b$th OFDM block is spatially precoded with $\wbf_l \in \mathbb{C}^{\Nt \times 1}$ and modulated as 
\begin{equation} \label{eq:si}
\sbf_b[k] = \frac{1}{\sqrt{\Nsc}} \sum_{l=0}^{\Nsc-1} \wbf_l x_{b,l} e^{\jrm 2 \pi \frac{lk}{\Nfft}}
\end{equation}  
for $-\Ncp \leq k \leq (\Nfft-1)$, and $\sbf_b[k] = 0$ otherwise; $\Nt$ is the number of transmitter antennas. 

OFDM blocks are then consecutively transmitted in the signal 
\begin{equation}
\sbf[k] =  \sum_{b} \sbf_b[k-b(\Ncp +\Nfft)].
\end{equation}   

\subsubsection{Channel}
The signal is conveyed to the receiver via a time-domain multi-antenna CIR of $L$ taps, satisfying
\begin{equation}
\label{eq:CIR}
\hbf[k]=\sum_{p=0}^{L-1} \hbf_p  \delta_{k,p}
\end{equation}
where $\hbf_p \in \mathbb{C}^{ \Nt \times 1}$ contains the channel path coefficients at delay $p$ (in samples). The entries of $\hbf_p $ are assumed to be i.i.d. and follow a zero-mean complex normal distribution 
with variance 
$E_p$ which is the energy of the $p$th tap. The total channel energy is written as $\alpha_L^2 = \sum_{p=0}^{L-1}  E_p$, and its discrete Fourier transform (DFT) at the $i$th subcarrier  is 
\begin{equation}
\label{eq:hhat}
\hcal_i = \sum_{k=0}^{\Nfft-1} \hbf[k] e^{-\jrm 2 \pi \frac{i k}{\Nfft}} =  \sum_{p=0}^{L-1} \hbf_p e^{-\jrm 2 \pi \frac{i p}{\Nfft}}. 
\end{equation}

We will assume that the maximum delay of the channel  is less than an OFDM symbol length, i.e. $L\leq \Nfft$, and the starting demodulation time at the receiver is synchronized with the first channel tap, such that the interference in the demodulated symbol, say $ \sbf_0$, is only caused by the previous block $\sbf_{-1} $. 

\subsubsection{Reception}   
The received signal is   
\begin{equation}
\label{eq:r[k]}
r[k]   = \sum_{m=0}^{L-1} \hbf^\Trm[m] \sbf[k-m] +z[k]
\end{equation}
where $z[k] $ 
is a zero-mean complex additive white Gaussian noise (AWGN) with variance $\sigma_{z}^2$.  
After CP removal, 
the received signal is then demodulated by FFT which gives as a demodulated
symbol on the $i$th subcarrier 
\begin{align}
y_i  &=  
\frac{\sqrt{\Nsc}}{\Nfft} \sum_{k=0}^{\Nfft-1} r[k] e^{-\jrm2 \pi \frac{ik}{\Nfft}} \nonumber \\
&=
\hcal^{\Trm}_{0,i,i}\wbf_i x_{0,i}  \label{eq:y[i]} \\ 
&\quad \quad  \quad + \underbrace{\sum_{\substack{l=0\\l\neq i}}^{\Nsc-1} \hcal^{\Trm}_{0,l,i}\wbf_l x_{0,l}}_\text{ICI} 
 + \underbrace{\sum_{\substack{l=0\\ ~}}^{\Nsc-1} \hcal^{\Trm}_{-1,l,i}\wbf_l x_{-1,l}}_\text{ISI} + n_i \nonumber
\end{align} 
where  $n_i= \frac{\sqrt{\Nsc}}{\Nfft} \sum_{k=0}^{\Nfft-1} z[k] e^{-\jrm2 \pi \frac{ik}{\Nfft}}$ is the post-processed AWGN with variance $\sigma^2_n = \frac{\Nsc}{\Nfft} \sigma_z^2$.  
The desired and interference OFDM channels $\hcal_{0,l,i}$ and $\hcal_{-1,l,i}$ can all be written as a DFT of a weighted/windowed CIR~\cite{PitavalArxiv21}. 
The desired signal multi-antenna channel is
\begin{equation}
\hcal_{0,i,i} =  \sum_{m=0}^{L-1} c[m] \hbf[m] e^{-\jrm 2 \pi \frac{im}{\Nfft}},
\label{eq:Hoii}
\end{equation}
the ICI multi-antenna channel coefficient from the $l\neq i$ subcarrier is 
\begin{equation}
\hcal_{0,l,i} =  \sum_{m=\Ncp}^{L-1} \tilde{c}_{l,i}[m] \hbf[m] e^{-\jrm 2 \pi \frac{lm}{\Nfft}},
\label{eq:Holi}
\end{equation}
and the ISI multi-antenna channel coefficient from the $l$th subcarrier is
\footnote{Precisely, 
$\hcal_{-1,i,i}=e^{\jrm 2 \pi \frac{i \Ncp}{\Nfft} }  \sum_{m=\Ncp}^{L-1}  (1-c[m]) \hbf[m] e^{-\jrm 2 \pi \frac{im}{\Nfft}} 
$ 
and 
$\hcal_{-1,l,i}=-e^{\jrm 2 \pi \frac{l \Ncp}{\Nfft} }  \sum_{m=\Ncp}^{L-1}  \tilde{c}_{l,i}[m]  \hbf[m] e^{-\jrm 2 \pi \frac{lm}{\Nfft}} 
$ for $l\neq i$,  
which can be compactly combined as~\eqref{eq:Hmli}. This reduction  as functions of~\eqref{eq:hhat},~\eqref{eq:Hoii} and~\eqref{eq:Holi} is 
because here the channel is causal w.r.t. the starting demodulation time, i.e., ISI comes only from the previous OFDM block. See~\eqref{eq:Gbii} and~\eqref{eq:Gbli} for comparison.  
} 
\begin{equation}
\label{eq:Hmli}
\hcal_{-1,l,i} = e^{\jrm 2 \pi \frac{l \Ncp}{\Nfft} } \left(\delta_{l,i} \hcal_i - \hcal_{0,l,i}\right) ,
\end{equation} 
with the weight functions as defined as below
\begin{equation}
c[m]= \begin{cases} \frac{\Nfft+m}{\Nfft} & -\Nfft \leq m \leq 0 \\
 1 &   0\leq m \leq \Ncp\\
\frac{\Nfft-(m-\Ncp)}{\Nfft} &    \Ncp \leq m \leq \Nfft+\Ncp\\
0 & \text{otherwise}
\end{cases}, \label{eq:cbias}
\end{equation}
and for $ l\neq i$. 
\begin{equation}
\tilde{c}_{l,i}[m] = \begin{cases}  
\frac{1-e^{\jrm 2 \pi \frac{m(l-i)}{\Nfft}}}{\Nfft \big(1-e^{\jrm 2 \pi \frac{(l-i)}{\Nfft}} \big)} & -\Nfft \leq m \leq 0 \\
  \frac{e^{\jrm 2 \pi \frac{(m-\Ncp)(l-i)}{\Nfft}}-1}{\Nfft \big(1-e^{\jrm 2 \pi \frac{(l-i)}{\Nfft}} \big)} & \Ncp \leq m \leq \Nfft+\Ncp \\
	0 & \text{otherwise}
\end{cases}. \label{eq:cbiasComplex}
\end{equation}

In order to write later more compact SINR expressions, we  extend the interference weight function to the case  $ l= i$ by
\begin{equation}\tilde{c}_{i,i}[m] = \frac{1-c[m]}{\sqrt{2}}\end{equation}
which will correspond to half the power of the  ISI from the demodulated subcarrier on itself. 

The simplifications above of the OFDM channels $\hcal_{0,l,i}$ and $\hcal_{-1,l,i}$  as single sum expressions mainly follow from classical OFDM derivations~\cite{PitavalArxiv21}, see~\cite{PhamTVT17,PhamPhD}  where derivation details are carried out for a causal channel (w.r.t. the stating demodulation time) and also earlier works such as~\cite{Viterbo96,Seoane97,Steendam99}. Writing~\eqref{eq:Hoii},~\eqref{eq:Holi} and~\eqref{eq:Hmli} as such directly simplifies greatly consecutive SINR derivations. 
Extensions of $c[m]$ and $\tilde{c}_{l,i}[m]$ to negative delays are provided here for later use in this paper. 

\subsubsection{SINR and Achievable Rate} 

Accordingly, the SINR on the $i$th subcarrier  with ${\rm SNR} = P/\sigma^2_n$ is 
\begin{multline}
\label{eq:SINRi}
{\rm SINR}_i = \\ 
\frac{ |\hcal^{\Trm}_{0,i,i} \wbf_i|^2}{|\wbf_l^\Trm\left(\hcal_i-\hcal_{0,i,i}\right)|^2 + \sum_{\substack{l=0\\l\neq i}}^{\Nsc-1}  2|\hcal^{\Trm}_{0,l,i}\wbf_l|^2 + 1/{\rm SNR}}.
\end{multline} 

Treating the interference as noise, an upper bound on an achievable rate for the given channel realization is
expressed in [bps/Hz]\footnote{An OFDM block duration is $T=\frac{\Nfft + \Ncp}{\Delta_f \Nfft}$ [s] and its bandwidth $B=\Delta_f \Nsc$ [Hz], thus the factor $\frac{1}{TB}=\frac{\Nfft}{(\Nfft+\Ncp)\Nsc}$.} by  
\begin{equation}
 \label{eq:rateTF}
R({\rm SNR}) = \frac{\Nfft}{(\Nfft+\Ncp)\Nsc} \sum_{i=0}^{\Nsc-1} \log_2(1+{\rm SINR}_i). 
\end{equation} 
This follows by treating the subcarriers as parallel channels, and bounding the mutual information of each 
to the one of a Gaussian additive noise channel with known noise variance, i.e. conditioned also on the knowledge of  the effective interference channel gains,  
see e.g.,~\cite{CaireTWC18}. The rate~\eqref{eq:rateTF} is thus an achievable rate under the assumption of perfect knowledge of channel state information at the receiver.

\subsection{Conventional Frequency-Selective Precoding} 
Multi-antenna precoding strategies and analyses are typically  based on the following type of frequency-flat channel equation 
\begin{equation}
y_i  =  \hcal^{\Trm}_i \wbf_i x_{0,i} + n_i
\end{equation}
which corresponds to a single-subcarrier transmission~\eqref{eq:y[i]} with sufficient CP, i.e. $L\leq\Ncp$, and where   the subcarrier index is often omitted for simplicity~\cite{sesia2011lte, Marzetta,HoydisJSAC13}.  
Multi-antenna precoding is thus typically applied individually over the subcarriers,  inherently frequency-selective,  and   constructed according to the  frequency response of the CIR $\hcal_i$. 
Accordingly, the conventional MRT precoding~\cite{LoMRT} is  
\begin{equation}
\label{eq:F-precoding}
\wbf_i = \frac{1}{\omega_i} \hcal_i^* 
\end{equation}
where  $\omega_i = \|\hcal_i\| $. 
While this MRT precoder follows from the reduction of the system model when  $L\leq \Ncp $, it was also  used as such in~\cite{aminjavaheri2017ofdm,NsengiyumvaMT}  for the case $L > \Ncp$  of an insufficient CP. In this paper, we will also consider precoding as~\eqref{eq:F-precoding} irreverently of the CP length and simply refer to it as \emph{F-precoding}. 
 
Since with~\eqref{eq:F-precoding}\footnote{By law of large number $\frac{\hbf_p^\Hrm \hbf_q}{\Nt} \to E_p\delta_{p,q} $, 
then $(\frac{\hcal^\Trm_i \wbf_i}{\sqrt{\Nt}})^2 = \frac{\|\hcal_i\|^2}{\Nt}= \sum_{p=0}^{L-1} \sum_{q=0}^{L-1} \frac{\hbf_p^\Hrm \hbf_q}{\Nt} e^{\jrm 2 \pi \frac{i(p-q)}{\Nfft}^2}  \to \sum_{p=0}^{L-1} E_p = \alpha_L^2$.}
 $\frac{\hcal^\Trm_i \wbf_i}{\sqrt{\Nt}} \to  \alpha_L$ as $\Nt \to \infty $ for all $i$, 
the effective channel of each subcarrier is asymptotically tending to the same AWGN channel: $y_i\sim  \sqrt{\Nt} \alpha_L x_{0,i} + n_i$.  In other words,  the frequency-selectivity (or fluctuation) of the overall effective channel is disappearing, i.e., it becomes frequency-flat. It is then tempting to conclude by time-frequency duality  that the overall multi-carrier transmission reduces asymptotically to a single-tap channel, and as a consequence, that the CP becomes unnecessary as $\Nt \to \infty$. 
This is not exactly correct because  the time-dispersion of the overall effective channel remains inside the CP as $\Nt \to \infty$, eventhough it disappears over the rest of the OFDM block if the CP is long enough.   
Defining the  single-antenna OFDM signal $s[k] =  \sum_{b} s_b[k-b(\Ncp +\Nfft)]$ with $s_b[k]= \frac{1}{\sqrt{\Nsc}} \sum_{l=0}^{\Nsc-1} x_{b,l} e^{\jrm 2 \pi \frac{lk}{\Nfft}}$ for $-\Ncp \leq k\leq\Nfft-1$ and $s_b[k]=0$ otherwise, we formalize this observation explicitly in the following whose derivation details can be found in Appendix~\ref{App:Lemma1}.

\begin{Lem} \label{lem:F-precoding}
With conventional F-precoding~\eqref{eq:F-precoding}, the transmission~\eqref{eq:r[k]} for the full OFDM block, $-\Ncp \leq k \leq  \Nfft-1$, is tending to a time-varying single-antenna two-tap channel, as $\Nt\to \infty$,
\begin{equation}
 \frac{r[k]}{\sqrt{\Nt}} \to  \frac{\beta_k^2}{\alpha_L}  s[k] + \frac{\alpha_L^2-\beta_k^2}{\alpha_L} s[k-\Nfft]
\end{equation}
where $\beta_k^2  = \sum_{m=0}^{ \min(k+\Ncp,L-1)} E_m$; which further reduces  to a  single-tap flat channel  only for the indices  $ L-\Ncp \leq k \leq \Nfft-1$: 
\begin{equation} \frac{r[k]}{\sqrt{\Nt}} \to \alpha_L s[k].
\end{equation} 
\end{Lem}

It follows that with F-precoding, the condition $L \leq \Ncp $ is still  required as $\Nt \to \infty$ to be able to demodulate a full OFDM symbol without ICI/ISI.

\subsection{Time-Frequency Selective Precoding} 

With $\Nt \to \infty$, a multi-antenna channel behaves asymptotically  as a spatial filter since  only its inner product with identical quantities remains significant, i.e. a component $\vbf \in \mathbb{C}^{\Nt\times1}$ in the transmitted multi-antenna signal  will be coherently combined  at the receiver if collinear with some channel paths, i.e., $\frac{\hbf_p^\Hrm\vbf }{\Nt} \to  \alpha E_p$ if $\vbf = \alpha \hbf_{p}$; otherwise it will be averaged out and be totally attenuated, as $\frac{ \hbf_{p}^\Hrm \vbf }{\Nt} \to 0$ for any $\vbf$ independent of $\hbf_{p}$.  
So in order to remove ISI/ICI, one should avoid to incorporate inside the precoder  components that correspond to certain channel paths that exceed a desired delay range. 
As the subspace spanned by the channel taps inside the CP, $\mathtt{span}\{\hbf_0,\ldots,\hbf_{\Ncp}\}$, is asymptotically orthogonal to the subspace spanned by the channel taps outside the CP, $\mathtt{span}\{\hbf_{\Ncp+1},\ldots,\hbf_{L-1}\}$, any precoder in   $\mathtt{span} \{\hbf_0,\ldots,\hbf_{\Ncp}\}$, i.e. any  linear combination of the form $\sum_{p=0}^{\Ncp}\lambda_p\hbf_p$, will asymptotically cancel out the ISI/ICI in the large antenna regime. This may nevertheless not be optimal  as channel taps with delay larger than the CP also contribute to the useful signal power, and potentially more than they contribute to the interference power~\cite{batariere2004cyclic}. Indeed, the useful channel~\eqref{eq:Hoii}  is a linear combination of all paths $\{\hbf_0,\ldots,\hbf_{L-1}\}$ while the interference channels~\eqref{eq:Holi} and~\eqref{eq:Hmli} are only combinations of the paths escaping the CP $\{\hbf_{\Ncp+1},\ldots,\hbf_{L-1}\}$.

Accordingly, we consider a truncated CIR with truncation threshold $\Th \leq L$ as 
\begin{equation}
\hbf^{(\Th)}[k] = \sum_{p=0}^{\Th-1} \hbf_p \delta_{k,p}, 
\end{equation} 
whose frequency response    
\begin{equation}
\hcal_i(\Th) =\sum_{k=0}^{\Nfft-1} \hbf^{(\Th)}[k] e^{-\jrm 2 \pi \frac{i k}{\Nfft}} =  \sum_{p=0}^{\Th-1} \hbf_p e^{-\jrm 2 \pi \frac{ip}{\Nfft}}
\end
{equation} 
can be used to derive conventional precoding methods.    
The MRT precoder on the $i$th subcarrier  $\wbf_i \in \mathbb{C}^{ \Nt \times 1}$  adapted to the truncated channel  is  therefore
\begin{eqnarray}
\label{eq:TrPrecoder} 
\wbf_i(\Th) &=& \frac{\hcal_i^*(\Th)}{\|\hcal_i (\Th)\|} 
=\frac{1}{\omega_i(\Th)} \sum_{p=0}^{\Th-1} \hbf_p^* e^{\jrm 2 \pi \frac{i p}{\Nfft}}
\end{eqnarray}
where   $\omega_i(\Th) = \| \hcal_i(\Th)\|$ is the normalization constant. We will simply refer to~\eqref{eq:TrPrecoder} as \emph{TF-precoding}. If $\Th=L$, the CIR is not truncated and TF-precoding falls back to F-precoding. 

TF-precoding does not limit itself to single-user MRT and can easily be  extended to other common linear precoders. We will nevertheless mainly focus on this  case as MRT is the main building block of other linear precoders that would lead to similar observations. 
A more detailed discussion for multi-user ZF precoding is for example provided  in Section~\ref{Sec:MU}. 

\section{Time-Reversal Filtering}
An alternative approach to pre-IFFT multi-antenna precoding is  post-IFFT time-reversal (TR) filtering,   as considered for example in~\cite{aminjavaheri2017ofdm} and similarly in~\cite{pitarokoilis2012optimality} for a single carrier system without any guard interval.  
TR-filtering with OFDM, and TEQ in general, shapes the signal after modulation which may create challenges for downlink adoption compared to reusing the legacy precoded-subcarrier  modulation structure. 
Nevertheless, this method without CP is asymptotically optimal (see Section IV-B) and thus provides an interesting benchmark. In the finite-antenna regime, it is less clear if TR-filtering would benefit of a CP, so we will  consider in general a TR system also with a CP for a fair comparison.  

\subsection{Transmission}
In this case, the multi-antenna OFDM signal is $\tilde{\sbf}[m] = [\tilde{s}_1[m],\ldots, \tilde{s}_{\Nt}[m]]^T $ obtained from a single OFDM signal 
convolved with a bank of matched-filters, i.e., for antenna $t$
\begin{equation}
\tilde{s}_t[m] = (h_t^{\rm TR} * s)[m] 
\end{equation}
where $h_t^{\rm TR}[n] = h_t^*[-n]/ \tilde{\omega}$ such that $\hbf[n] = [h_1[n], \cdots,h_{\Nt}[n] ]^\Trm$, and $s[n]$ is as defined before Lemma 1. The normalization constant is   $\tilde{\omega}^2= \sum_{p=0}^{L-1} \|\hbf[p]\|^2 $ to satisfy $ \mathbb{E}_{\{x_{b,l}\}}\left[\|\tilde{\sbf}[m] \|^2\right] = 1$. 
The received signal is 
\begin{eqnarray}
\tilde{r}[k] &=& \sum_{m=0}^{L-1} \hbf^\Trm[m]\tilde{\sbf}[k-m] +z[k] \\ 
&=& 
(g*s)[k] + z[k]
\end{eqnarray}
where the effective channel is 
\begin{equation}
g[n] = \sum_{t=1}^{\Nt} (h_t*h_t^{\rm TR})[n] = \sum_{m=0}^{L-1} \hbf^{\Hrm}[m-n]\hbf[m]/\tilde{\omega}.
\end{equation}

As a result, the transmission reduces to a single-antenna OFDM system with non-causal channel $g[n]$ and the demodulated signal is a function of the previous and next OFDM blocks. The  channel $g[n]$ has a pulse-like shape, symmetric around $n=0$, and the interference can be minimized by adding a delay of $\Delta = \Ncp/2$ as a starting demodulation time  so that the bulk of its impulse response is centered inside the CP. The effective channel then  becomes $g'[n]=g[n-\Delta ]$. 
Therefore, after demodulation of the samples with indices $\Delta  \leq k \leq \Delta+\Nfft-1 $, the received signal for the $i$th subcarrier is

\begin{align}
\tilde{y}_i &=\gcal_{0,i,i}x_{0,i} \label{eq:y[i]TR} 
\\
&\quad + \underbrace{\sum_{\substack{l=0\\l\neq i}}^{\Nsc-1} \gcal_{0,l,i} x_{0,l}}_\text{ICI} 
 + \underbrace{\sum_{\substack{l=0\\ ~}}^{\Nsc-1} \left( \gcal_{-1,l,i} x_{-1,l} + \gcal_{1,l,i} x_{1,l}\right) }_\text{ISI}+ n_i,\nonumber
\end{align}
where the desired signal channel is
\begin{equation} \label{eq:G0ii}
\gcal_{0,i,i} = e^{-\jrm 2 \pi \frac{i \Delta}{\Nfft}} \!\!\!\sum_{m=-L+1}^{L-1}\!\!\!\!\! c[m+\Delta] g[m] e^{-\jrm 2 \pi \frac{im}{\Nfft}},
\end{equation}
the ICI channel coefficient from the $l\neq i$ subcarrier is 
\begin{equation} \label{eq:G0li}
\gcal_{0,l,i} =  e^{-\jrm 2 \pi \frac{l \Delta}{\Nfft}} \!\!\!\sum_{m=-L+1}^{L-1} \!\!\!\!\! \tilde{c}_{l,i}[m+\Delta] g[m] e^{-\jrm 2 \pi \frac{lm}{\Nfft}},
\end{equation}
the ISI channel from the $i$th subcarrier of the ($b=\pm1$)-block is  
\begin{equation} \label{eq:Gbii} 
\gcal_{b,i,i} =   e^{-\jrm 2 \pi \frac{i (b\Ncp+\Delta)}{\Nfft}}\mkern-40mu\sum_{m\in b\times \{-L+1,\cdots,0\}} \mkern-40mu(1-c[m+\Delta]) g[m] e^{-\jrm 2 \pi \frac{im}{\Nfft}}, 
\end{equation} 
and the ISI channel from the $l\neq i$th subcarrier  of the ($b=\pm1$)-block is
\begin{equation} \label{eq:Gbli}
\gcal_{b,l,i} =  - e^{-\jrm 2 \pi \frac{l (b\Ncp+\Delta)}{\Nfft}}\mkern-40mu \sum_{m\in  b\times\{-L+1,\cdots,0\}} \mkern-40mu\tilde{c}_{l,i}[m+\Delta] g[m] e^{-\jrm 2 \pi \frac{lm}{\Nfft}}, 
\end{equation} 
with $c[m]$ and $\tilde{c}_{l,i}$ as given in~\eqref{eq:cbias} and~\eqref{eq:cbiasComplex}, respectively. 

Accordingly, the SINR on the $i$th subcarrier  is 
\begin{multline}
\label{eq:SINRiTR}
{\rm \widetilde{SINR}}_i= \\ 
\frac{ |\gcal_{0,i,i}|^2}{\sum_{\substack{l=0\\l\neq i}}^{\Nsc-1}  |\gcal_{0,l,i}|^2   +
\sum_{l=0}^{\Nsc-1}  \left(|\gcal_{-1,l,i}|^2 +|\gcal_{1,l,i}|^2\right) 
+ 1/{\rm SNR}}, 
\end{multline}
and the corresponding transmission rate is 
\begin{equation} \label{eq:rateTR}
\tilde{R}({\rm SNR}) = \frac{\Nfft}{(\Nfft+\Ncp)\Nsc} \sum_{i=0}^{\Nsc-1} \log_2(1+{\rm \widetilde{SINR}}_i). 
\end{equation} 

\subsection{Relationship and Differences between TR-filtering and TF-Precoding}

TR-filtering provides a  theoretical benchmark since it provides asymptotically the optimum performance. However, a post-IFFT technique corresponds to a change of modulation structure compared to a pre-IFFT technique, and as a result the implementation of  TR-filtering may raise more practical concerns than TF-precoding, notably in the downlink. We discuss here some correspondences and differences between the two methods. 
\paragraph{A partial time/frequency implementation equivalence} 
TR-filtering and TF-precoding are partially connected by Fourier transforms of each other. In general, implementations of post-IFFT filtering and pre-IFFT precoding are equivalent but only for the output samples which do not have any OFDM symbol overlap at the filter's input~\cite{FP-OFDM}. This would be the case here between TR-filtering and TF-precoding if one ignores the normalization constants. 
In fact, part of the transmitted signal of TF-precoding if $\Th \leq \Ncp $ could equivalently  be obtained by a form of TR-filtering where the filters would be truncated versions of the $\{h_t^{\rm TR}\}$. 
The prefixes in  these two implementations would however still differ.  The filtered-CP would depend on the previous OFDM block, i.e., the  CP after filtering is not a cyclic extension anymore, while by construction the CP with TF-precoding is. 

\paragraph{TR-filtering spreads the signal}   TR-filtering like any filtering inherently creates a tail on the transmitted signal, increasing the transmission time and thus overhead~\cite{FP-OFDM}. In turn, a potential benefit is that, some, but uncontrolled, out-of-band reduction could be obtained. 

\paragraph{TR-filtering distorts the in-band spectrum} Normalization techniques are different for the two methods. In TF-precoding, each subcarrier can be individually normalized, while in TR-filtering the normalization is only on the overall signal and thus the same for all subcarriers. It results that TR-filtering will modify  the signal spectrum  according to the channel spectrum and thus will increase in-band fluctuation.  This feature may not be desirable in the downlink for most practical systems which typically have in-band distortion constraints via error-vector magnitude (EVM)~\cite{3GPPTS36.101} or spectral flatness~\cite{LitePoint} requirements. 

\paragraph{Frequency-division multiplexing} For frequency-division multiplexing of different users,  with TR-filtering each user would need its own OFDM modulation to be filtered according to its specific channel  before combining. With TF-precoding, different sub-bands could be precoded independently before passing through the same OFDM modulation.

\section{Asymptotic Rate Analysis} 

We will focus on the asymptotic regime $\Nt \to \infty$ with ${\rm SNR} \to 0$  such that the operational SNR, ${\rm SNR_{op}} = \Nt {\rm SNR}$,  is fixed. A similar asymptotic regime is considered e.g. in~\cite{NgoTCOM13} for energy efficiency, but this  
  is also relevant to approximate the performance under a fixed quality of service  as provided by a finite-size  constellation. 
Indeed with a fixed constellation size,  the SNR needed to reach e.g. a certain SER level is shifting in the low SNR-regime with the precoding gain $10 \log_{10} \Nt$  [dB] as $\Nt$ increases. Meanwhile and as we are investigating interference mitigation methods, we will primarily focus on relatively high-SNR regimes (i.e. ${\rm SNR_{op}}$ high) in which interference is the limiting factor.

\subsection{Asymptotic Rate of TF-Precoding} 
We have the following result (see Appendix~\ref{App:Prop1}). 
\begin{Prop} \label{Prop:SINR}
With TF-precoding~\eqref{eq:TrPrecoder}, the SINR defined in~\eqref{eq:SINRi} is tending  to 
\begin{multline} \label{eq:SINRprop}
 {\rm SINR}_i^{\infty}(\Th) = \\ \frac{  \left( \sum_{p=0}^{\Th-1}  c[p] E_p \right)^2 }{  
\sum_{l=0}^{\Nsc-1}   2\left| \sum_{p=\Ncp+1}^{\Th-1}  \tilde{c}_{l,i}[p] E_p \right|^2
+\alpha^2_{\Th} {\rm SNR_{op}^{-1}} }
\end{multline}
as $N_t \to \infty$ with ${\rm SNR_{op}}=\Nt {\rm  SNR} >0 $ fixed, and where  $\alpha_{\Th}^2 = \sum_{p=0}^{\Th-1} E_p$. 
\end{Prop}

The corresponding asymptotic rate is
 \begin{multline}
R^\infty({\rm SNR_{op}} ; \Th) = \\ \frac{\Nfft}{(\Nfft+\Ncp)\Nsc} \sum_{i=0}^{\Nsc-1} \log_2(1+{\rm SINR}_i^\infty(\Th)).   \label{eq:rateTFinf}
\end{multline}
The special case $\Th = L$ gives the asymptotic rate of the conventional F-precoding~\eqref{eq:F-precoding}. 

The asymptotic form ${\rm SINR}_i^{\infty}(\Th) $ is dependent of the subcarrier index $i$. This is because not all subcarriers in the IFFT are occupied, and as a result, subcarriers in the middle of the band receive more interference from neighboring subcarriers, and as a consequence have a lower SINR than the edge subcarriers.  
If all subcarriers are occupied ($\Nsc=\Nfft$), without cyclic prefix ($\Ncp=0$), and  without  time-selectivity in the precoder ($\Th=L$), we recover the expression derived in~\cite{aminjavaheri2017ofdm}. Note also that compared to~\cite{aminjavaheri2017ofdm,PitavalGlobecom18}, we have simplified from the beginning the system model in~\eqref{eq:y[i]} such that Prop.~\ref{Prop:SINR} is almost direct from the SINR definition provided in~\eqref{eq:SINRi}.

Here however, unlike with conventional F-precoding, we observe that with an appropriate truncation threshold in the precoder design, the system is not necessarily interference-limited. 
\begin{Cor}
If $\Th \leq \Ncp+1$ then  ${\rm SINR}_i^{\infty}(\Th) = \alpha_{\Th}^2 {\rm SNR}_{\rm op}$, i.e., ISI and ICI vanish as $N_t \to \infty$.
\end{Cor}

One way to apprehend this result is that  TF-precoding with $\Th \leq \Ncp+1$  asymptotically transforms the linear convolution of the signal with the channel to a circular convolution with only channel taps inside the CP. To see this, we can write the received signal as $r[k]   = \sum_{m=0}^{\Ncp} \hbf^\Trm[m] \sbf[k-m] + \sum_{n=\Ncp+1}^{L-1} \hbf^\Trm[n] \sbf[k-n]$, splitting between a first term which is equivalent to  a circular convolution of the signal with the channel paths inside the CP for $0\leq k\leq \Nfft-1$, while the other is a linear convolution with the tail of the channel escaping the CP. Accordingly, the demodulating symbol~\eqref{eq:y[i]} can be written as    
$y_i = \hcal^\Trm_{{\rm CP},i}\wbf_i x_{0,i} + y_{{\rm tail},i} + n_i$ where the circular-convolution channel component $\hcal_{{\rm CP},i} = \sum_{m=0}^{\Ncp}  \hbf[m] e^{-\jrm 2 \pi \frac{im}{\Nfft}}$ is the DFT of only the channel taps inside the CP, while $ y_{{\rm tail},i} $ contains the linear-convolution components from the channel taps outside the CP. Precisely, we have $ y_{{\rm tail},i}  = \hcal^\Trm_{{\rm tail},i}\wbf_i x_{0,i} + \sum_{\substack{l=0\\l\neq i}}^{\Nsc-1} \hcal^\Trm_{0,l,i}\wbf_l x_{0,l}  + \sum_{\substack{l=0\\ ~}}^{\Nsc-1} \hcal^\Trm_{-1,l,i}\wbf_l x_{-1,l}$ with 
$ \hcal_{{\rm tail},i}= \hcal_{0,i,i}- \hcal_{{\rm CP},i}$, and in which each effective channel is a weighted DFT of only the channel paths outside the CP.  

By selecting $\Th =\Ncp+1$, $\wbf_i=\frac{ \hcal^*_{{\rm CP},i}}{\|\hcal_{{\rm CP},i}\|}$ and the TF-precoder is made to match exactly this circular-convolution component. At the same time, this enables also to asymptotically remove the linear-convolution channel components as all effective channels in $ y_{{\rm tail},i} $ depend only of independent channel taps outside the CP, leading to  $ y_{{\rm tail},i} /\sqrt{\Nt} \approx 0 $. Therefore, the transmission reduces asymptotically to a circular convolution as 
\begin{equation*}
\frac{y_i}{\sqrt{\Nt}} \approx \frac{\hcal^\Trm_{{\rm CP},i}\wbf_i}{\sqrt{\Nt}}  x_{0,i}
=\frac{\|\hcal_{{\rm CP},i}\|}{\sqrt{\Nt}} x_{0,i} \to \alpha_{\Th}  x_{0,i}.
\end{equation*}

Meanwhile, the term $\alpha_{\Th}$ represents a loss factor compared to the maximum possible received SNR with full multipath combining gain, which would then be $ \alpha_{L}^2 {\rm SNR}_{\rm op}$. Remark also that $ y_{{\rm tail},i} $ is not purely interference and contains a desired signal contribution. A threshold larger than the CP might thus increases the SINRs, and as a result a trade-off appears between interference mitigation and multipath combining gain for rate maximization. 

\subsection{Asymptotic Rate of TR-Filtering}
TR-filering asymptotically provides the maximum possible received SNR. 
With $\hbf^{\Hrm}[m]\hbf[p]/\Nt \to E_m \delta_{m,p}$  as $\Nt \to \infty$,  we verify that $\tilde{\omega}/\sqrt{\Nt} \to \alpha_L$ and the effective channel of TR-filtering is asymptotically tending to a single-tap (frequency-flat) channel,  
$g[n]/\sqrt{\Nt} \to \alpha_L \delta_n$. So, independently of the fact that the transmitted signal is an OFDM one,  the SINR of such transmission is asymptotically tending to 
${\rm \widetilde{SINR}}_i\to \alpha_L^2 {\rm SNR_{op}}$ 
as $\Nt \to \infty$. Moreover, since asymptotically the delay spread of the effective channel is zero, no CP is needed, leading  to the optimal rate 
\begin{equation}
\label{eq:rateTRinf} 
\tilde{R}^{\infty}({\rm SNR_{op}}) = \log_2(1+\alpha_L^2 {\rm SNR_{op}}).
\end{equation}
Compared to TF-precoding, TR-filtering without CP in the asymptotic regime  neither suffers of interference nor of  CP overhead, and thus provides an ultimate performance upper bound. 

\subsection{Optimization and Comparison}
We now numerically optimize the precoder for the standard Extended Typical Urban (ETU) channel model and  3GPP parameters of  $\Nsc=600$ subcarriers separated by $\Delta_f = 60$ kHz, $\Nfft = 2048$, and a \emph{normal CP}  of $\Ncp=144\approx 7\% \Nfft $. 
We consider that the CP length is a semi-static system parameter which cannot be changed as dynamically as  the precoders, and so we consider a precoder optimization given a CP length. In addition, if CP adaptation is possible, it is likely to be from a look-up table with limited values so that it can be efficiently signaled. We  therefore limit the numerical analysis to the CP lengths available in 5G NR~\cite{3GPPTS38.211}, and consider in addition no CP for TR-filtering. 

\begin{figure}[t]
\vspace{-0.2cm} 
\includegraphics[width=.5\textwidth]{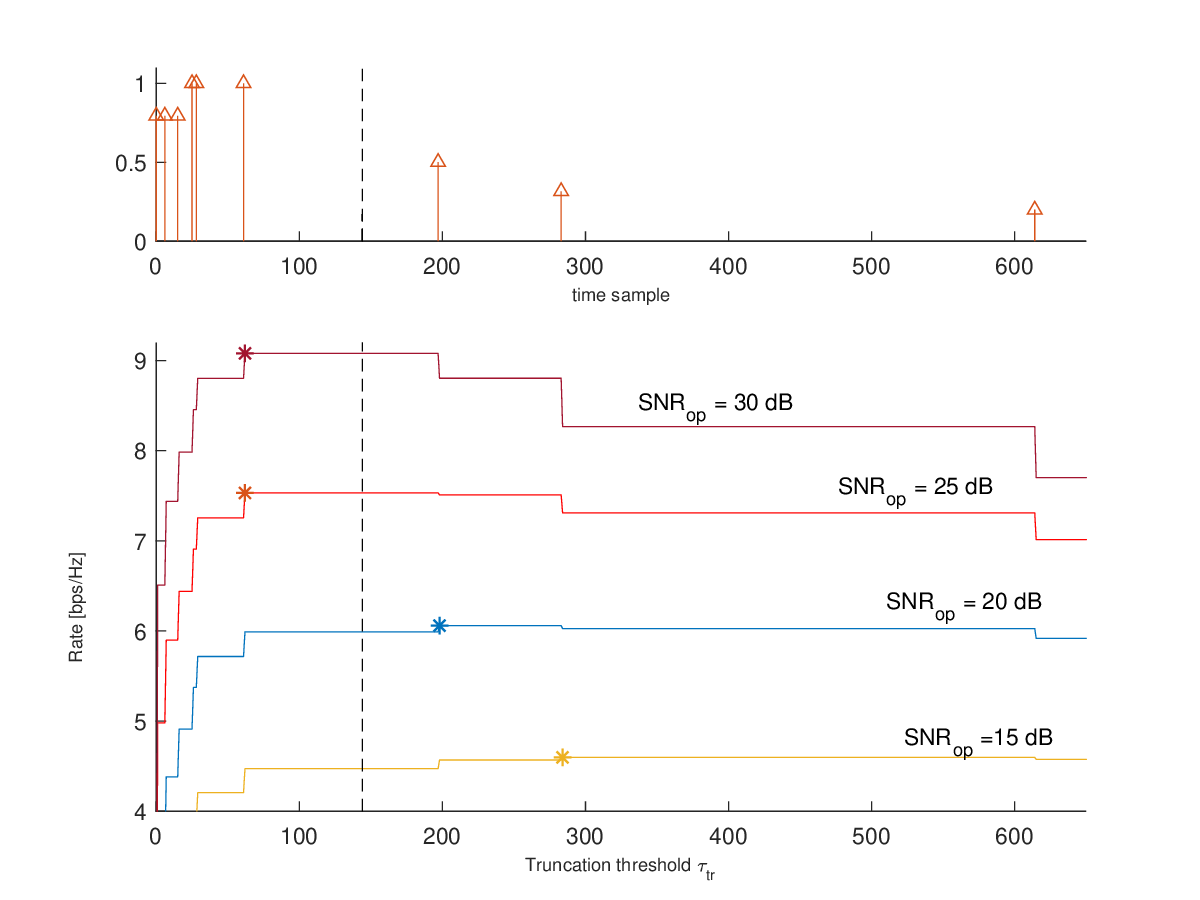}
\vspace{-0.6cm}
	\caption{Asymptotic ($\Nt \to \infty$) achievable rate of TF-precoding~\eqref{eq:rateTFinf} as a function of the truncation threshold $\Th$. Top: ETU channel's PDP; Bottom: Asymptotic rate $R^\infty({\rm SNR_{op}} ; \Th)$. The vertical dashed line indicates the CP length.\label{fig:Optimization}}
\vspace{-0.0cm}
\end{figure}

\subsubsection{Optimized  Truncation Threshold of TF-precoding at a given ${\rm SNR_{op}}$}     
For each channel realization, one could search for a threshold in~\eqref{eq:TrPrecoder} that maximizes the rate~\eqref{eq:rateTF} as
$\tau_{\max}({\rm SNR}) = \arg \max_{\Th} R({\rm SNR}; \Th).$ 
This optimization can be simplified in the asymptotic regime 
by maximizing instead the rate~\eqref{eq:rateTFinf}  as given by 
$\tau_{\max}^\infty ({\rm SNR_{op}}) = \arg \max_{\Th} R^{\infty}({\rm SNR_{op}}; \Th).$  
This asymptotic optimization only depends of the system parameters,  the operational SNR,  and the average channel energy of the taps (which can be approximated from a single channel acquisition in the large antenna regime as $\| \hbf_p\|^2/N_t \to E_p$  with $N_t\to \infty $).  
The optimum threshold, which may not be unique, can be found by exhaustive search.   
The search only needs  to be started from $\tau_{\Th}= \Ncp+1$ as any threshold inside the CP would be suboptimal.
Then as $\tau_{\Th}$ increases, the rate expression will only change when a non-zero channel tap is included in the precoder, and thus the size of the search follows from the number of non-zero taps rather than time samples.   
As ${\rm SNR_{op}} \to 0$, the interference term becomes negligible in~\eqref{eq:SINRprop} and  $\tau_{\max}^\infty \to L$. On the other hand as ${\rm SNR_{op}} \to \infty$ the interference dominates and typically $\tau_{\max}^\infty \to \Ncp+1$.  

The normalized power delay profile (PDP) of the ETU channel is displayed on the upper part of Fig.~\ref{fig:Optimization}. 
As it can be seen, with a 60 kHz SCS, 3 channel taps are escaping the normal CP length. 
The lower part of Fig.~\ref{fig:Optimization} shows the asymptotic rate~\eqref{eq:rateTFinf} 
of  TF-precoding  as a function of the truncation threshold $\Th$ and for different operational SNRs. For each SNR, the smallest threshold value maximizing the rate is indicated by a star-symbol. 
In the high-SNR regime, ${\rm SNR_{op}} \gtrsim 25$ dB, one verifies that the rate is maximized by precoding only according to the channel paths inside the CP and thus here $\tau_{\max}^\infty=\Ncp+1$ maximizes the rate. For ${\rm SNR_{op}} = 20 $ and $15  $ dB, the rate is maximized with a $\Ncp+1 < \tau_{\max}^\infty < L$, selecting some, but not all, paths outside the CP. 
We observe that selection of the optimum truncation threshold has only a significant impact in the high-SNR regime, i.e., when the threshold is set to match the CP length as $\Th = \Ncp+1$. 
Similar results and conclusions can be found in~\cite{PitavalGlobecom18} for the TDL-C channel~\cite{3GPPTR38.900}, which is a denser version of the ETU channel with 24 taps. TDL-C has a tunable delay scaling selected in~\cite{PitavalGlobecom18} to 2 $\mu $s, i.e. twice larger than the maximum recommended value in~\cite{3GPPTR38.900},  in order to better highlight the trade-off between interference cancellation and multipath-combining gain in the precoding design. In general, this trade-off is exacerbated  if some strong channel paths are arriving shortly after the end of the CP. Such scenario would equivalently occur for a TDL-C channel with 1 $\mu$s and a twice larger SCS of $120$ kHz.   

\subsubsection{Rate Comparison}

\begin{figure}[t]
\vspace{-0.2cm}
\includegraphics[width=.5\textwidth]{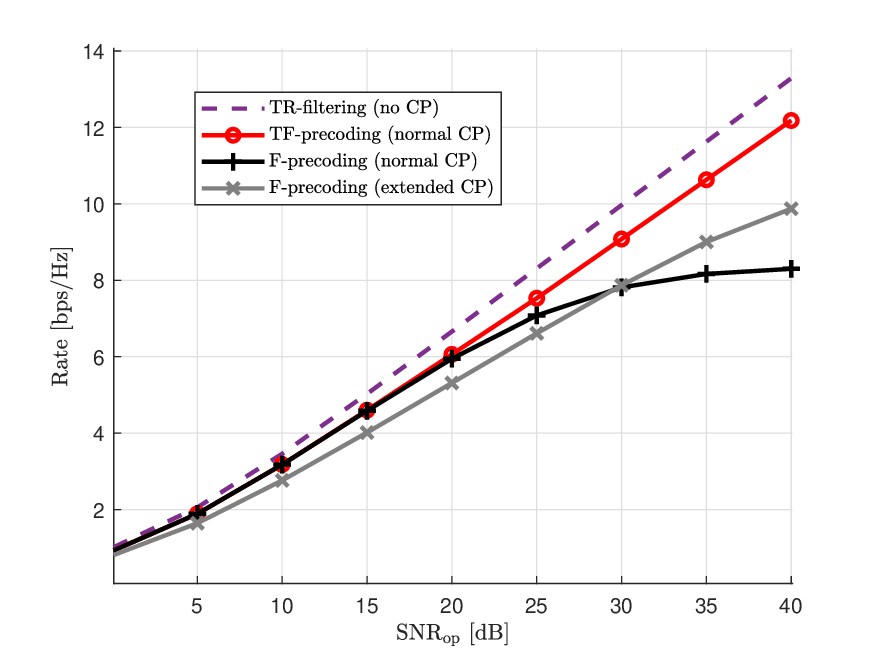}
\vspace{-0.6cm}
	\caption{Optimized asymptotic rate $R^\infty({\rm SNR_{op}} ; \tau_{\max}^\infty ({\rm SNR_{op}} ) )$  of TF-precoding compared to F-precoding with normal and extended CP, and TR-filtering without CP. \label{fig:CapaOptvsSNRop}    } 
\vspace{-0.0cm}
\end{figure}

Fig.~\ref{fig:CapaOptvsSNRop} compares the rate of  TF-precoding, F-precoding, and TR-filtering according~\eqref{eq:rateTFinf} and~\eqref{eq:rateTRinf}. The channel and system parameters are  as before with also the consideration of an \emph{extended CP}  of $\Ncp=512 = 25\% \Nfft  $ as defined in~\cite{3GPPTS38.211}.  
For TF-precoding, the truncation threshold is selected as $\tau_{\max}^\infty(\leq 10 {\rm dB}) = L$, $\tau_{\max}^\infty( 15 {\rm dB}) =  284$, $\tau_{\max}^\infty( 20 {\rm dB}) = 198$ and   $\tau_{\max}^\infty(\geq 25 {\rm dB})=\Ncp+1$,  following the optimization above.  
As expected, TF-precoding always outperforms F-precoding  
and provides notable improvements in the high-SNR region where F-precoding is interference-limited. Using an extended CP with F-precoding converts interference to useful power and thus improves the SINR in the high-SNR regime. However, this solution suffers of a high rate penalty from CP overhead. 
TF-precoding  enables to not increase the CP overhead and still removes or mitigates interference. 
TR-filtering, having neither any CP overhead  nor interference,  provides asymptotically the best performance. 

\section{Finite-size Rate Analysis }

This section provides average rate approximations for finite numbers of antennas. It will show that TF-precoding converges faster to its asymptotic rate than TR-filtering with and without CP, providing analytical support to the 
link-level simulation results in Section VI. 

\subsection{Approximated Rate Expressions}

To get a simple but instructive analysis, we follow a similar methodology as in~\cite{LimTWC15}. 
The average rate can be written in the form $ \expect{\log_2 \left(1+\frac{\Scal}{\Ical+{\rm SNR}^{-1}}\right)}$.  
Inside this expression, the signal and interference power tend to deterministic quantities as $\frac{1}{\Nt}\Scal\to K_{\Scal}$ and   $\frac{1}{\Nt}\Ical \to K_\Ical$ where $0 \leq K_\Ical,\, K_\Scal<\infty$, c.f. the proof of Prop. 1. 
So with ${\rm Var } \left[\frac{1}{\Nt}\Scal\right] \to 0$ and ${\rm Var } \left[\frac{1}{\Nt}\Ical\right] \to 0$, 
we can approximate   
$\frac{1}{\Nt}\Scal \approx \frac{1}{\Nt} \expect{\Scal} \to K_{\Scal}$,  $\frac{1}{\Nt}\Ical \approx \frac{1}{\Nt} \expect{\Ical} \to K_{\Ical}$.  
With ${\rm SNR_{op}} < \infty  $,  the SINR is then well approximated by 
$\frac{\Scal}{\Ical+{\rm SNR}^{-1}} = \frac{\frac{1}{\Nt} \Scal}{\frac{1}{\Nt} \Ical+(\Nt{\rm SNR})^{-1}} \approx \frac{\frac{1}{\Nt} \expect{\Scal}}{\frac{1}{\Nt} \expect{\Ical}+{\rm SNR_{op}^{-1}}}$,  
from which follows the  rate approximation
\begin{align}
&\expect{\log_2 \left(1+\frac{\Scal}{\Ical+{\rm SNR}^{-1}}\right)} \nonumber \\ 
&\mkern150mu \approx  \log_2\left(1+\frac{\expect{\Scal}}{\expect{\Ical}+{\rm SNR}^{-1}}\right) \label{eq:RateApprox}
\\ &\mkern150mu \to  \log_2\left(1+\frac{K_{\Scal}}{K_{\Ical}+{\rm SNR_{op}^{-1}}}\right). 
\end{align} 
Note that if, e.g. for some $\Th> \Ncp$, the interference  does not vanish as $\Nt \to \infty$  
then  
this rate approximation is also well-defined for any SNR, without requiring $\Nt{\rm SNR} \to {\rm SNR_{op} } <\infty $.

We use a similar argument to separately average out  the precoder normalization in the SINR expression; from a system point of view this corresponds to approximate the downlink and uplink performance to be the same, which is asymptotically the case.  
We also remark that defining instead the SINR  directly as the \emph{average-signal-power} to \emph{average-interference-power}-plus-noise ratio as in the right-hand side of~\eqref{eq:RateApprox} 
has often been used in the literature as a convenient OFDM design parameter, see e.g.,~\cite{Steendam99,batariere2004cyclic,Mostofi06,aminjavaheri2017ofdm}. 

\subsubsection{TF-precoding} It follows, 
 see details in Appendix~\ref{App:FS_TF}, that the average rate of TF-precoding is asymptotically approximated as  
\begin{equation} \label{eq:rateTFapprox}
\expect{R({\rm SNR_{op}}; \Th) } \approx \frac{\Nfft}{(\Nfft+\Ncp)\Nsc} \sum_{i=0}^{\Nsc-1} \log_2(1+\Gamma_i(\Th)) 
\end{equation} 
where $\Gamma_i(\Th)$ is given in \eqref{eq:SINRexpect} at the top of the next page.  
One can easily verify that $\Gamma_i(\Th)  \to {\rm SINR}_i^{\infty}(\Th) $ and so this provides indeed a tight approximation   as $\Nt \to \infty$. 
\begin{figure*}[t]
\begin{equation} \label{eq:SINRexpect}
 \Gamma_i(\Th) = 
\frac{  \Nt\left( \sum_{p=0}^{\Th-1} c[p] E_p \right)^2 +  \alpha_{\Th}^2 \sum_{m=0}^{L-1} c[m]^2 E_m}
{  
\sum_{l=0}^{\Nsc-1}  2\left(   \Nt \left| \sum_{p=\Ncp+1}^{\Th-1} \tilde{c}_{l,i}[p] E_p \right|^2 +  \alpha_{\Th}^2 \sum_{m=\Ncp+1}^{L-1} | \tilde{c}_{l,i}[m]|^2 E_m \right)
+ \Nt \alpha_{\Th}^2  {\rm SNR_{op}^{-1}} }
\end{equation}
\hrulefill
\vspace*{4pt}
\end{figure*}

\subsubsection{TR-filtering}
With similar derivations, see Appendix~\ref{App:FS_TR},  the average rate of TR-filtering is asymptotically approximated as  
\begin{equation} \label{eq:rateTRapprox}
\expect{\tilde{R}({\rm SNR_{op}}) } \approx \frac{\Nfft}{(\Nfft+\Ncp)\Nsc}\sum_{i=0}^{\Nsc-1} \log_2(1+\tilde{\Gamma}_i) 
\end{equation} 
where 
\begin{equation}
\label{eq:SINRiTRapprox}
\tilde{\Gamma}_i = \frac{\Nt \alpha_L^4 + \sum_{m=-L+1}^{L-1} c[m+\Delta]^2 \rho_m  }
{\sum_{l=0}^{\Nsc-1}  \sum_{m=-L+1}^{L-1} 2 |\tilde{c}_{l,i}[m+\Delta]|^2   \rho_m + \Nt \alpha_L^2  {\rm SNR_{op}^{-1}}}
\end{equation}
with $\rho_m =  \sum_{n=0}^{L-1}  E_n E_{n-m}$. It can also be verified that $\tilde{\Gamma}_i \to {\rm \widetilde{SINR}}_i^{\infty}$ and so it is indeed a tight approximation  as $\Nt \to \infty$. 

In the special case there is no CP ($\Ncp=\Delta = 0$) and all subcarriers are occupied ($\Nsc= \Nfft$), the expression is further simplified as 
\begin{equation}
\label{eq:SINRiTRapproxNfft}
\tilde{\Gamma}_i = \frac{\Nt \alpha_L^4 + \sum_{m=-L+1}^{L-1} c[m]^2 \rho_m  }
{ \sum_{m=-L+1}^{L-1} (1-c[m]^2)\rho_m + \Nt \alpha_L^2  {\rm SNR_{op}^{-1}}}. 
\end{equation}
This last expression  with  $\alpha_L =1$ matches the expression in~\cite{aminjavaheri2017ofdm} while it has been derived differently.

\subsection{Verification and Comparison}

\begin{figure}
\vspace{-0.0cm}
\subfigure[TF-precoding \label{fig:FSapprox_TFprecoding}]{\includegraphics[width=.5\textwidth]{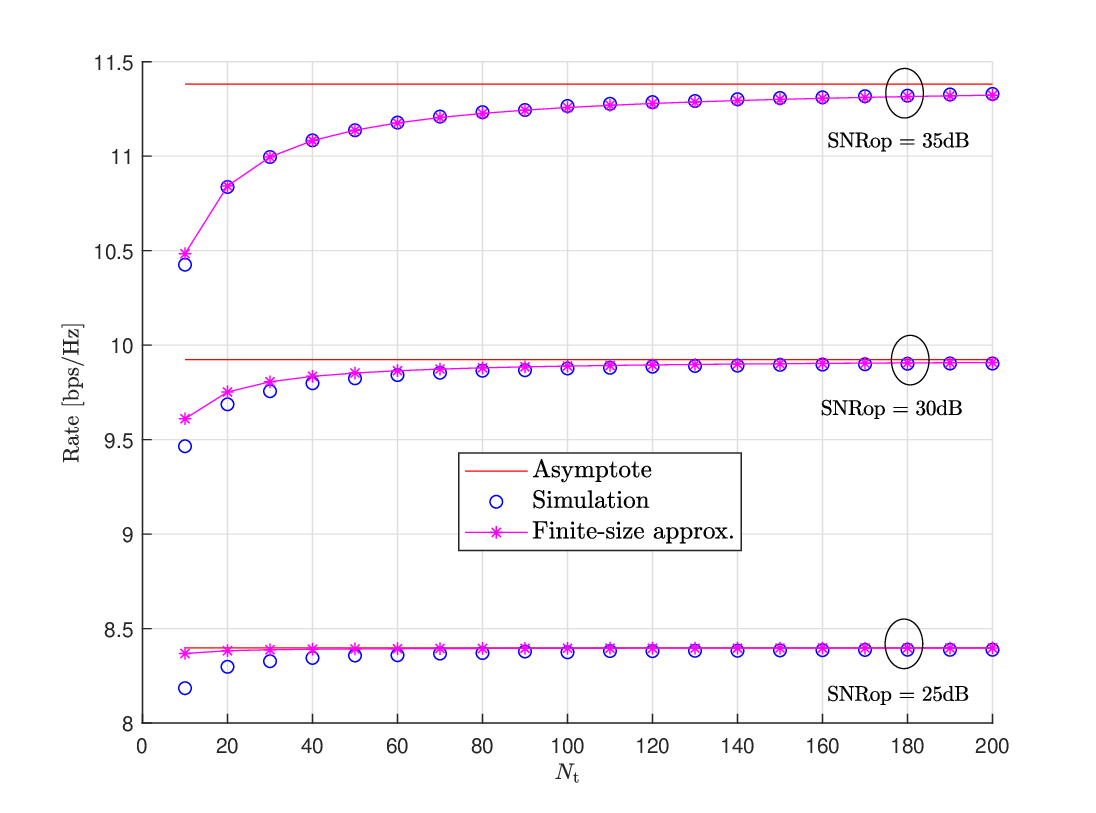}}
\subfigure[TR-filtering  \label{fig:FSapprox_TRfiltering}]{\includegraphics[width=.5\textwidth]{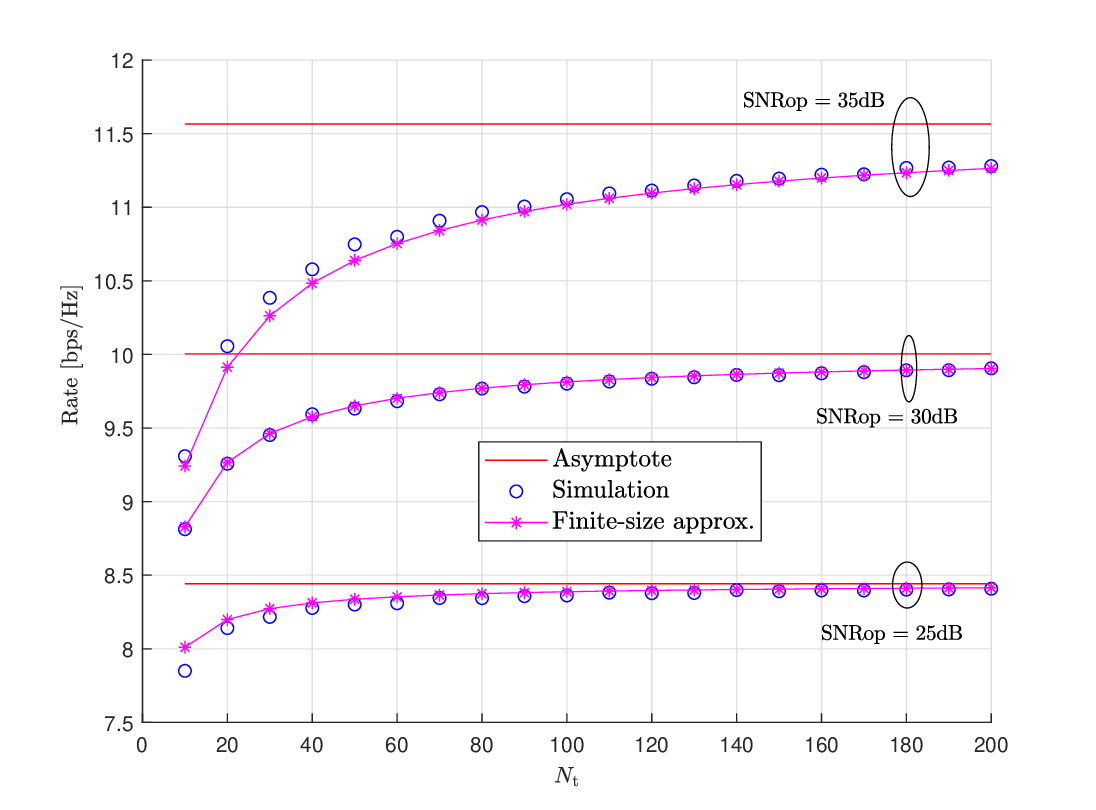}}
	\vspace{-0.1cm}
	\caption{Comparison of the rate approximations~\eqref{eq:rateTFapprox} and~\eqref{eq:rateTRapprox} to simulated averages of~\eqref{eq:rateTF} and~\eqref{eq:rateTR}, respectively, as well as their asymptotes~\eqref{eq:rateTFinf} and~\eqref{eq:rateTRinf}. The channel has an exponential-decaying PDP here. \label{fig:FSapprox_verification}}
 \vspace{-0.0cm}
\end{figure}

Using some small parameter values, $\Nfft=32$,  $\Nsc=12$, $\Ncp = 2$ and a fixed $\Th = 4$, we verify in  Fig.~\ref{fig:FSapprox_verification} the derived rate approximations~\eqref{eq:rateTFapprox} and~\eqref{eq:rateTRapprox} from $\Nt=10$ to $200$ compared to the simulated Monte-Carlo averages of~\eqref{eq:rateTF} and~\eqref{eq:rateTR}, as well as their convergence to the asymptotic rates~\eqref{eq:rateTFinf} and~\eqref{eq:rateTR}. Also for the purpose of verification, we use in this case a synthetic exponential-decaying PDP given by $E_p =  e^{-p}$ for $p=0,...,\Nfft-1$, so that the energy of all channel taps is non-zero for any possible delays. The approximations match well the simulations even with a small number of antennas. For both cases, it can be verified that the rate approximations tend to underestimate/overestimate the rate for some low/high ${\rm SNR}_{\rm op}$, such that these approximations are neither upper nor lower bounds. 
Even though TR-filtering has an higher asymptote, it can be observed that TF-precoding converges faster to its own asymptote. The TF-precoding rate curves start at $\Nt=10$ with already more than 90\% of the asymptote ($\approx$ 97, 95 and 92\% at ${\rm SNR_{op}} = 25,\, 30 $ and 35 dB, respectively), while TR-filtering reaches 90\% of the asymptotic rate only  at $\Nt \approx 10,\,  15$ and $30$ for  ${\rm SNR_{op}} = 25,\, 30$ and 35 dB, respectively.

\begin{figure}[t]
\centering
\vspace{-0.2cm}
\includegraphics[width=.5\textwidth]{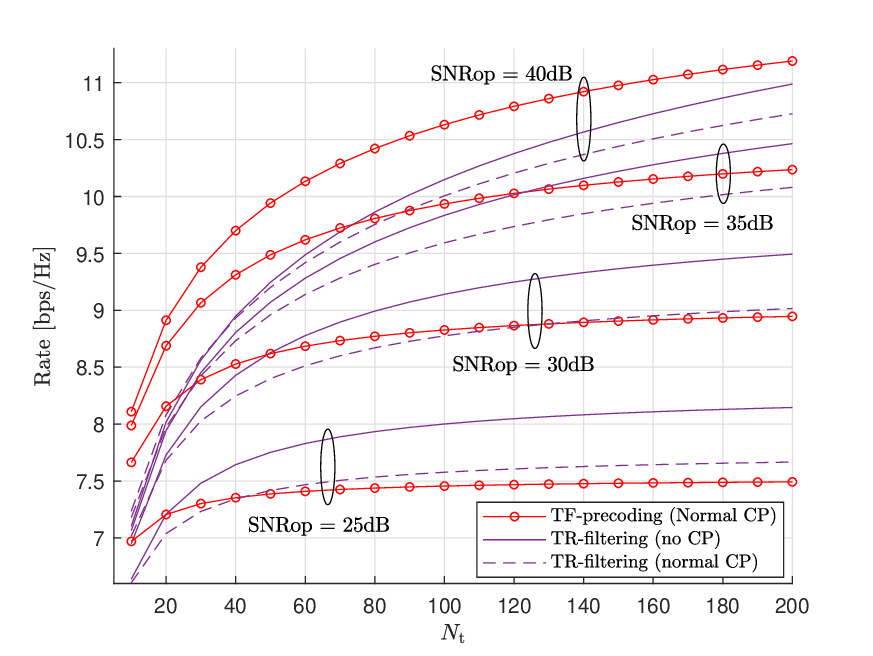}
\vspace{-0.6cm}
	\caption{Average rate comparison between TF-precoding and TR-filtering in ETU channel as a function of the number of antennas.\label{fig:TRTFcomparison}}
\vspace{-0.0cm}
\end{figure}

Similar curves are displayed on Fig.~\ref{fig:TRTFcomparison} but for the scenario of Fig.~\ref{fig:CapaOptvsSNRop} discussed previously, i.e., $\Nsc=600$, $\Nfft = 2048$ and $\Ncp=144 $ with ETU channel. Here, the truncation threshold in TF-precoding is adapted to the SNR according to the asymptotic optimization of the previous section. 
Using a normal CP with TR-filtering only marginally improves the rate for a  small number of antennas at high SNR as it slightly outperforms the case without CP only for $\Nt\leq 16,\,25$ and $35$ with ${\rm SNR_{op}} = 30,\, 35$ and 40 dB, respectively.
This means that it does not provide a subsequent SINR gain that would compensate the CP overheard. Remark that TR-filtering  doubles the channel delay-spread which was already longer than the normal CP length, leading thus to   
more non-negligible interfering taps  if only few antennas are used. 
Because the TF-precoding has to deal with less interference than TR-filtering, 
its rate for very small number of antennas is higher than TR-filtering both with and without CP.
It follows that in the high-SNR regime, TF-precoding can outperform TR-filtering for not-so-large antenna arrays. The two curves cross at  $\Nt \approx 20,\, 50,\, 120$ and 310, for  ${\rm SNR_{op}} = 25,\, 30,\, 35 $ and 40 dB,  respectively.  

\begin{figure}[t]
\centering
\vspace{-0.0cm}
\includegraphics[width=.5\textwidth]{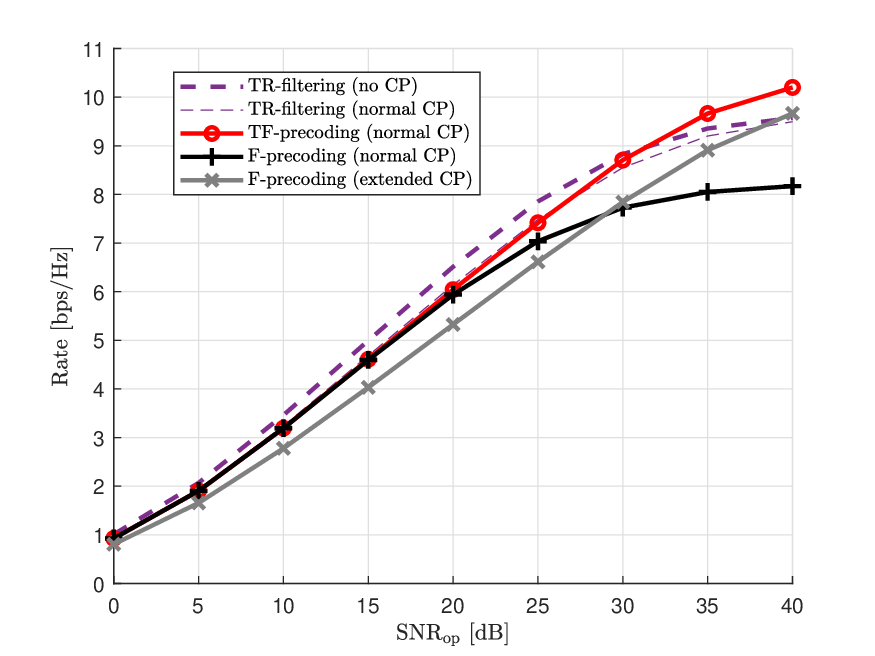}
\vspace{-0.6cm}
	\caption{Average rate comparison of TF-precoding, F-precoding, and TR-filtering with 64 antennas and ETU channel.\label{fig:TRTFcomparaison_SNRop}}
\vspace{-0.0cm}
\end{figure}

In Fig.~\ref{fig:TRTFcomparaison_SNRop}, the average rates as a function of ${\rm SNR_{op}} $ with 64 antennas  are compared for TF-precoding with normal CP, F-precoding with normal/extended CP, and TR-filtering with/without normal CP. Other parameters are as in Fig.~\ref{fig:CapaOptvsSNRop}. Compared to the asymptotic case,  TF-precoding and TR-filtering are now both interference-limited, but still outperform F-precoding with normal or extended CP. TF-precoding notably maintains relatively similar gains over F-precoding. TF-precoding and TR-filtering have now more comparable performance. TR-filtering without CP provides slightly better rate than TF-precoding in the low and middle SNR range, after which in the high-SNR regime,   TF-precoding starts to perform the best as it is less interference-limited. 

\section{Link-Level Simulations}

In this last section, we compare the different designs by SER and throughput link-level simulations with  finite size arrays of $\Nt= 64$ and $200$ antennas. We consider the transmission of i.i.d. 4, 16 and 64 quadrature amplitude modulation (QAM) symbols over block-fading Rayleigh channels with independent realizations every subframe of 14 OFDM blocks. Simulations are performed by direct implementation of all functions, including OFDM modulation, CP addition, time-domain channel propagation, CP removal and OFDM demodulation. 
Other simulation assumptions are as before, cf. Fig~\ref{fig:CapaOptvsSNRop}.  

Received symbols are equalized according to their effective channels given in~\eqref{eq:y[i]} and~\eqref{eq:y[i]TR}, i.e. $\hcal^\Trm_{0,i,i} \wbf_i$ and $\gcal_{0,i,i}$. We remark that if the effective channels would be instead assumed to be from the Fourier transform of the CIRs as it is the case with a sufficient CP, this would translate into a slight performance degradation, mainly for TR-filtering.  The received symbols are then detected by maximum-likelihood decoding. 

For TF-precoding, a unique truncation threshold  per constellation is used following the optimization of Section IV.  
Because the throughput from a finite constellation typically converges  to its maximum in a rather  small SNR region, the optimization  can indeed be  performed only for a unique SNR value where e.g. the constellation is expected to reach a sufficiently-low error rate. We select $\Th= 284$, $198$, and $\Ncp+1$, for 4, 16 and 64 QAM, respectively, targeting a good performance in the $10^{-3}$-$10^{-4}$ SER region.

\subsubsection{SER}
\begin{figure*}[t]
\vspace{-0.4cm}
\subfigure[SER with $\Nt=64$ \label{fig:SER_64}]{\includegraphics[width=.50\textwidth]{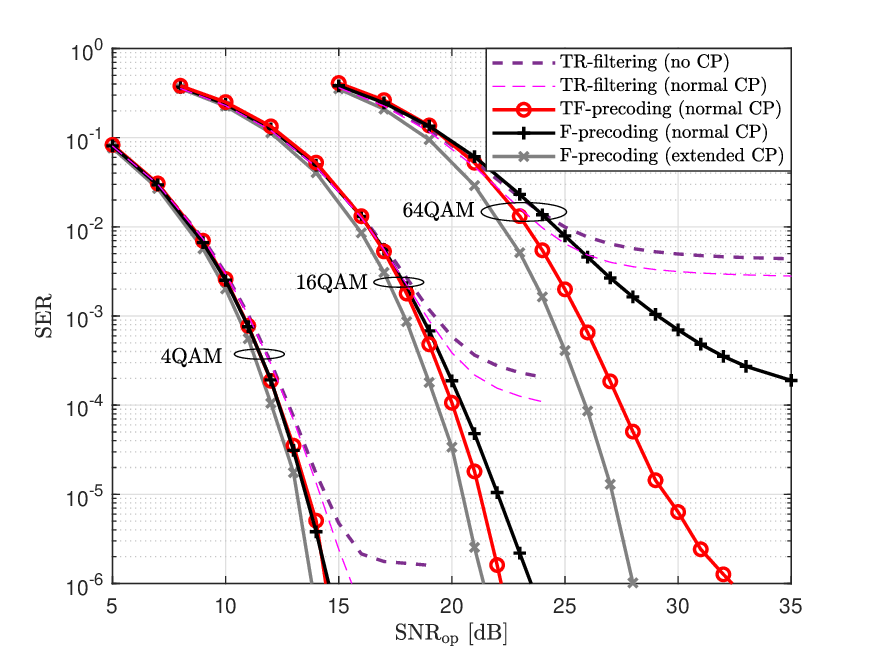}} 
\subfigure[SER with $\Nt=200$ \label{fig:SER_200}]{\includegraphics[width=.50\textwidth]{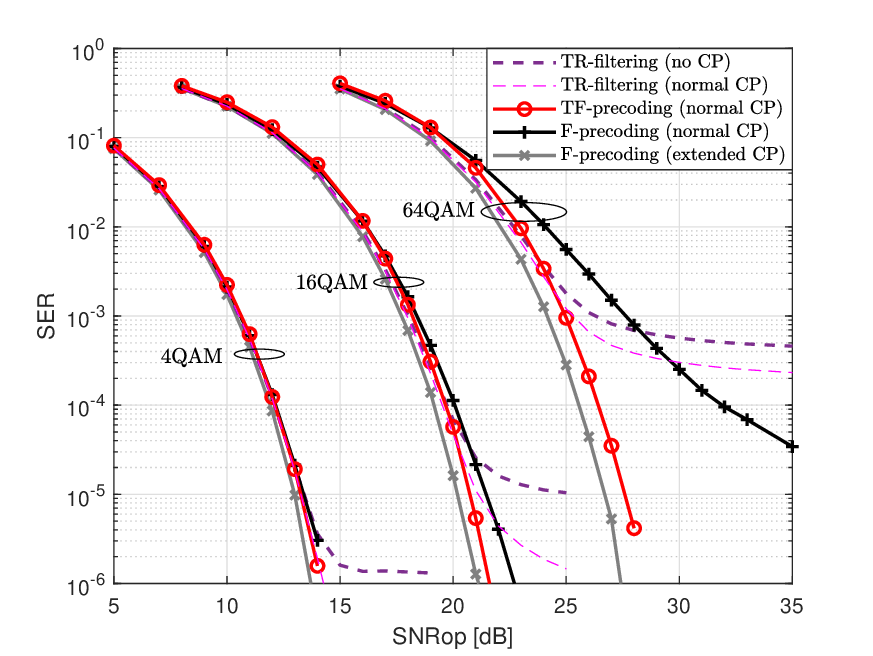}}
\subfigure[Throughput with $\Nt=64$  \label{fig:T_64}]{\includegraphics[width=.50\textwidth]{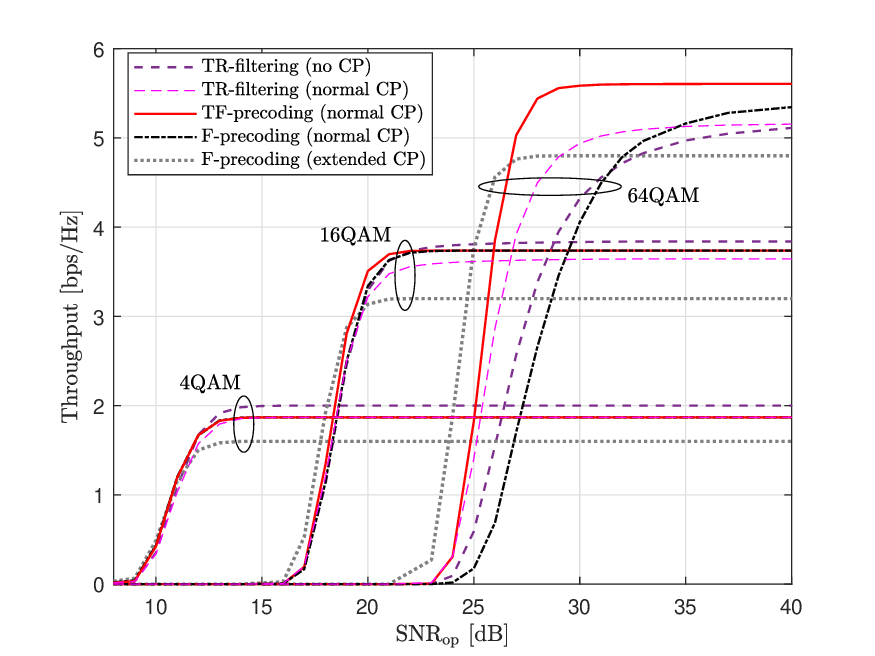}}
\subfigure[Throughput with $\Nt=200$  \label{fig:T_200}]{\includegraphics[width=.50\textwidth]{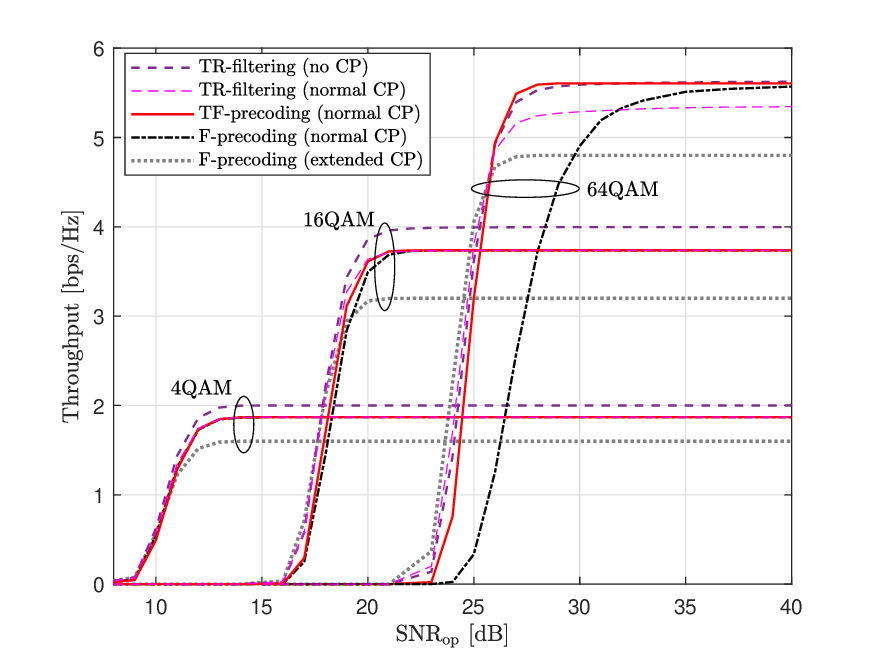}}
\vspace{-0.1cm}
	\caption{Link level simulations in ETU channel. Upper and lower parts are SERs and throughputs, respectively. Left and right parts are with $64$ and $200$ transmitter antennas, respectively. \label{fig:LLS}}
	\vspace{-0.0cm}
\end{figure*}

SERs are shown on the upper part of Fig.~\ref{fig:LLS}. F-precoding with an extended CP performs the best since it includes almost all the channel paths in its CP and thus does not suffer of interference. On the contrary, the performance of F-precoding with a normal CP is notably degraded with 64 QAM. In this case, TF-precoding can provide a subsequent SNR gain from F-precoding while maintaining a normal CP. TR-filtering is reaching an error floor in all cases and the level of this floor is lowered down when using a CP. This error floor is more significant with larger constellation sizes and smaller numbers of antennas. 
The performance of all schemes improves with a larger number of antennas, but only marginally for F-precoding with an extended CP since it barely has any ISI/ICI.  

\subsubsection{Throughput}

The SER comparison does not balance with the fact that the schemes have different CP lengths and thus operate at different rates. To have a fairer comparison, the lower part of Fig.~\ref{fig:LLS} displays the throughput of each design. 
The throughput is numerically obtained from the same simulations as
\begin{equation}
T =  \log_2(Q) \frac{\Nfft}{\Nfft+\Ncp}\left(1-{\rm BLER}\right) 
\end{equation}
where $Q$ is the modulation order and ${\rm BLER}$ is the simulated  error probability  of blocks of 
$\Nsc=600$ constellation symbols.  
The throughput is upper bounded by $\frac{\Nfft}{\Nfft+\Ncp}\log_2(Q) $ which is reached for a sufficiently high SNR, conditioned there are no SER error floor. More precisely, the maximum throughput is $\log_2(Q)$, $0.93 \log_2(Q)$ and $0.8 \log_2(Q)$, with zero CP, normal CP, and extended CP, respectively. 

As it can be seen from the figures, when the throughput converges to its maximum, the extension of the CP for F-precoding does incur a high rate loss; but it also provides a subsequent gain compared to a normal CP with 64QAM for a large SNR region, justifying its design. Meanwhile, TF-precoding provides a much higher rate with 64QAM since it maintains the CP length while also mitigates the interference. 
TR-filtering without CP performs the best with $\Nt= 64$ and 4QAM, and  $\Nt= 200$ and 4QAM and 16QAM, by reaching quickly the maximum rate of $\log_2(Q) $ without CP overhead. Even though TR-filtering without CP could  potentially converge to a higher maximum with 64QAM, its SER  floor is too high   to provide a satisfactory performance with  $\Nt= 64$, and starts only to  perform similarly to TF-precoding  with  $\Nt= 200$. Using a normal CP with TR-filtering improved only the former case of  $64$-QAM with $\Nt=64$  such that the throughput curve is converging faster to its maximum but this maximum remains  the same  than without CP.

\section{Discussions}
\subsection{Spatial Correlation}

\begin{figure*}[t]
\vspace{-0.2cm}
\subfigure[Rate simulations as in Fig.~\ref{fig:FSapprox_verification} with different $\rho$ \label{fig:RateSimuSC}]{\includegraphics[width=.48\textwidth]{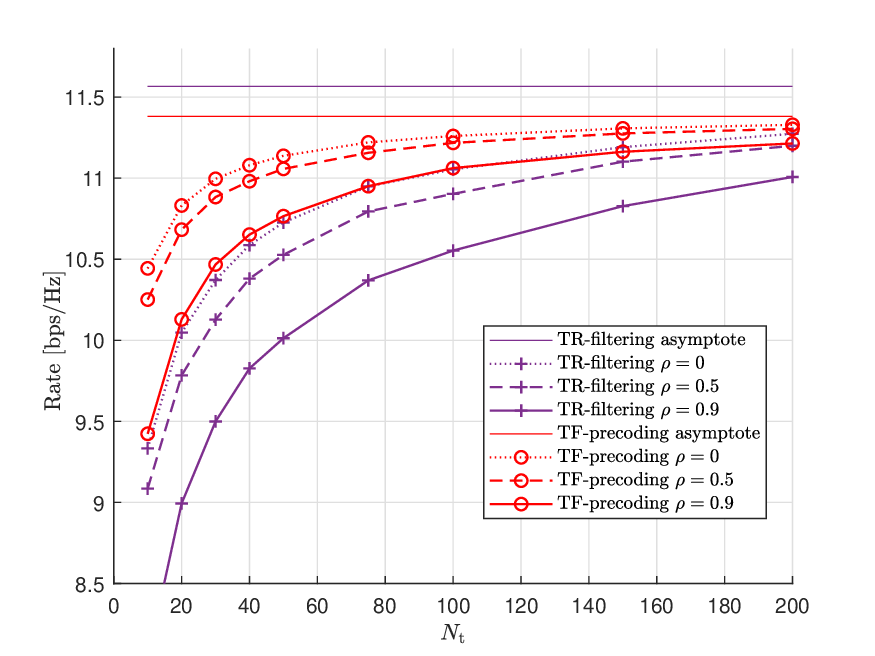}} 
\subfigure[Throughput with $\Nt=200$, 64 QAM, and $\rho =0.9$.  
\label{fig:T_SpatCorr}]{\includegraphics[width=.48\textwidth]{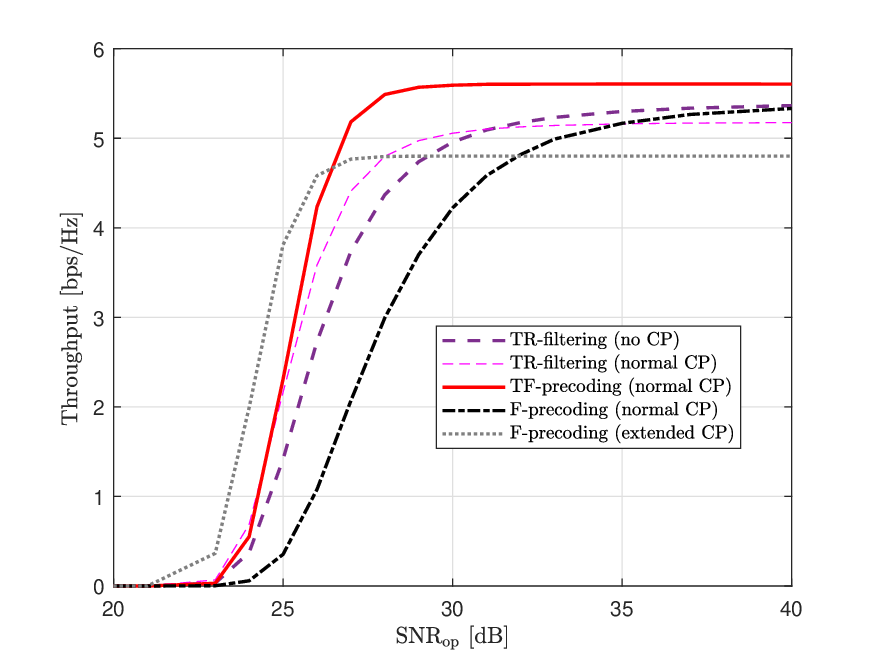}}
\vspace{-0.1cm}
	\caption{Simulations with spatial correlation.  Exponential correlation model with correlation factor $\rho$. \label{fig:SpatCorr}}
	\vspace{-0.0cm}
\end{figure*}

We briefly discuss  the impact of spatial correlation on the performance conclusions drawn from the provided i.i.d. channel model analysis. 
Indeed, practical multi-antenna transmissions  are likely to have some spatial correlation, nevertheless i.i.d. analysis of large-array systems has been shown to approximate fairly well the characteristics of realistic systems~\cite{RusekSigMag13,RusekTWC15}. 
As  typically done, the  analysis in this paper was based on the asymptotic orthogonality of random channel vectors and also  convergence of channel gains to their means; facts that  under the i.i.d. assumption directly follow from the law of large numbers.    
If  considering spatially-correlated Rayleigh fading instead,  
these properties would remain\footnote{A sufficient condition  is that the spectral norm of the spatial correlation matrices is
bounded, and the average channel gains remain strictly positive~\cite{HoydisJSAC13,BjornsonFTSP17}.} and thus asymptotic performance would be unchanged, however, convergence  is expected to be slowed down in the sense that more antennas would be required to achieve the same performance than with i.i.d. channels~\cite{HoydisJSAC13,BjornsonFTSP17}.  

To exemplify this, we consider the simple exponential spatial correlation model for a uniform linear array with correlation factor $\rho$
between adjacent antennas~\cite{Loyka}. Namely, the spatial  spatial correlation matrix of the $p$-tap is $\Rbf_p$ with entries $[\Rbf_p]_{k,l} = \rho^{|k-l|} {\rm exp}(\jrm(k-l)\phi_p)$ where $\phi_p$ is  the angle of departure.  
These angles are selected to be in the range $[-\frac{\pi}{3},\, \frac{\pi}{3}]$, uniformly distributed across different channel realizations.  

In Fig.~\ref{fig:RateSimuSC}, we reproduce the simulations of Fig.~\ref{fig:FSapprox_verification}  with ${\rm SNR_{op}}  = 35$ dB for $\rho= 0,\,0.5$ and $0.9$. The case $\rho= 0$ corresponds to an i.i.d. channel. As expected, introduction of spatial correlation slows down the convergence of the system to its asymptotic regime. We observe that the performance drop is higher with TR-filtering.    
In Fig.~\ref{fig:T_SpatCorr}, we then reproduce the simulations of Fig.~\ref{fig:T_200}   with 64 QAM and 
 $\rho= 0.9$. The performance of all schemes is degraded compared to Fig.~\ref{fig:T_200}, but this is insignificant for F-precoding with extended CP and small for TF-precoding, while it is rather large for F-precoding with normal CP and TR-filtering with/without CP.  
The level of degradation corresponds to the level of interference there is in each scheme.  
Again, the convergence to the asymptotic behavior, and thus the expected interference mitigation,  is slowed down so that the performance comparison here with spatial correlation and $\Nt= 200$  is  closer to Fig.~\ref{fig:T_64} with i.i.d. channel and $\Nt= 64$ antennas.  

More important for the analysis considered in this paper is the assumption of statistical independence between the channels taps. 
This is a standard assumption, in line with the wide-sense stationary uncorrelated scattering (WSSUS) model where signals (or clusters of  non-resolvable paths) arriving at different delays  are uncorrelated. Such model is typically consider valid for most wireless channels~\cite{Steendam99}.

\subsection{Time-windowed precoder:} 
\label{sec:TwinP}
A  possible refinement of the truncated precoder design discussed would be to insert some window function inside the precoder instead of a simple truncation (being itself actually a rectangular windowing).  
As revealed  by the formulation given in~\eqref{eq:y[i]}, a good choice for such window function could be the bias function $c[n]$ so that precoding would be  better matched to the desired signal channel with insufficient CP, $\hcal_{0,i,i}$. In fact, such design would slightly improve the performance over the conventional F-precoding in the high-SNR regime, but  would  nevertheless still be suboptimal if no truncation is used. 
As the SNR goes to infinity, we have seen that the optimal precoding strategy tends to  totally exclude paths outside the CP in order to guarantee a total interference cancellation. Also w.r.t. truncation optimization in the mid-range SNR regime, one can remark that $c[n]\approx 1$ for the paths outside but in the vicinity of the CP, and therefore the impact of inserting such windowing inside the precoder is negligible. 

\subsection{Multi-User Extension}
\label{Sec:MU} 
As shown in~\cite{aminjavaheri2017ofdm} for the case of no CP, multi-user interference  is asymptotically canceling out and 
does not play a role in the  SINR saturation that comes from  an insufficient CP. Therefore, the asymptotic rate analysis in Section IV along  with its truncation threshold optimization can directly be extended to the multi-user case. 

As a multi-user channel  is the concatenation  of several single-user channels, multi-user TF precoders can be obtained similarly from individual-user truncated CIRs. 
For a transmission to $\Nu$ users, each with a single receive antenna, the $\Nu \times \Nt$  multi-user CIR is $\Hbf[n]$
with Fourier transform $\Hcal_i$ on the  $i$th subcarrier.  
Individual truncation can be applied to each user channel. Namely, the channel of user $u$, $\hbf^\Trm_u[n]$ being the $u$th row of  $\Hbf[n]$,  is truncated at $\tau_u$, leading to the truncated CIR $\hbf_u^{(\tau_u)} [n]$.  
A truncated multi-user channel is then obtained as the concatenation of individual-user truncated channels as $\Hbf^{(\bm{\tau}_{\rm tr})}[n]=[\hbf_1^{(\tau_1)}[n],\ldots,\hbf_{\Nu}^{(\tau_{\Nu})}[n]]^\Trm$ with  $ \bm{\tau}_{\rm tr} = [\tau_1,\ldots,\tau_{\Nu}]^\Trm$; and its frequency-response at the $i$th subcarrier is denoted by   $\Hcal_i(\bm{\tau}_{\rm tr})$.  

\begin{figure}[t]
\centering
\vspace{-0.2cm}
\includegraphics[width=.5\textwidth]{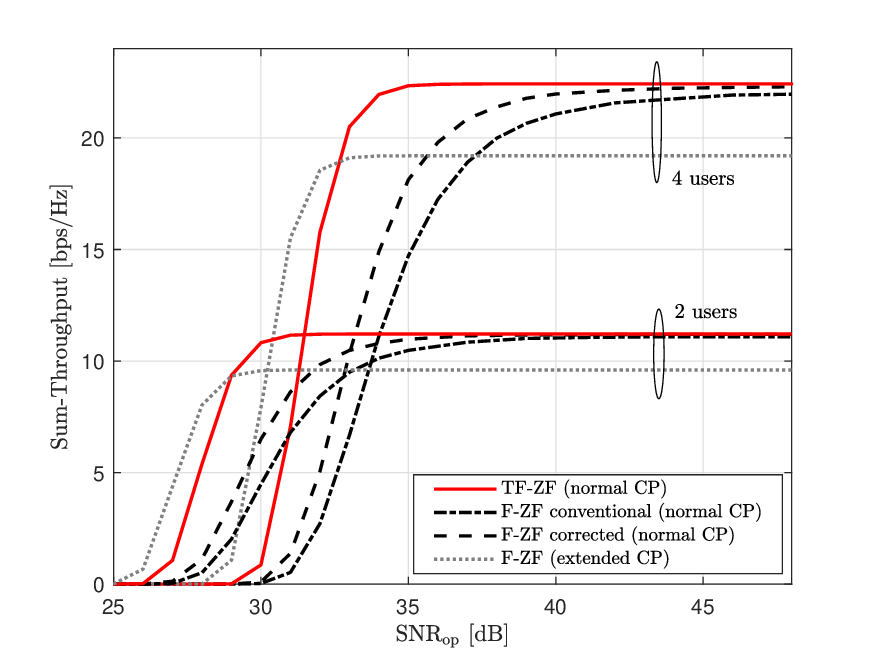}
\vspace{-0.6cm}
	\caption{Sum-throughput in multi-user scenarios using ZF precoders~\eqref{eq:TF-ZF},~\eqref{eq:F-ZF} and~\eqref{eq:F-ZF2}.  $\Nt=200$  antennas  and 64 QAM. \label{fig:MU}}
	\vspace{-0.0cm}
\end{figure}

From the obtained  truncated multi-user channel, conventional multi-user precoders can be computed to construct corresponding TF precoders. For example, omitting the normalization and power allocation among users,  a multi-user TF-MRT is thus obtained as $\Wbf_i^{\rm (TF-MRT)} = \Hcal^\Hrm _i(\bm{\tau}_{\rm tr})$. 
For ZF, it is first worth remarking that according to~\eqref{eq:y[i]} a ZF precoder with a proper matrix inversion should be of the form $ \Vbf (  \Hcal_{0,i,i}  \Vbf)^{-1}$ and not $ \Vbf (  \Hcal_{i}  \Vbf)^{-1}$ as the desired signal channel with an insufficient CP is actually  $ \Hcal_{0,i,i} =  \sum_{m=0}^{L-1} c[m] \Hbf[m] e^{-\jrm 2 \pi \frac{im}{\Nfft}}$ and not $\Hcal_{i} $. 
For this reason, the conventional F-ZF,  
\begin{equation} \label{eq:F-ZF}
\Wbf^{\rm (F-ZF)} =   \Hcal_{i}^\Hrm  (  \Hcal_{i}   \Hcal_{i}^\Hrm )^{-1}, 
\end{equation}
used as such in~\cite{aminjavaheri2017ofdm,NsengiyumvaMT} for an insufficient CP, is ill-defined. 
This might cause already significant
performance degradation compared to a corrected F-ZF with a proper matrix inversion as     
$\Hcal_{i}^\Hrm  (   \Hcal_{0,i,i}     \Hcal_{i}^\Hrm )^{-1}$, which can further be improved as discussed in Section VII-B to better match the desired signal channel by 
\begin{equation}\label{eq:F-ZF2}
\Wbf^{\rm (F-ZFc)}  =    \Hcal_{0,i,i}^\Hrm  (   \Hcal_{0,i,i}     \Hcal_{0,i,i}^\Hrm )^{-1}.
\end{equation}
Following the same observation, a TF-ZF is accordingly given by\footnote{As discussed in Section~\ref{sec:TwinP}, this could be modified to 
$\Wbf^{\rm (wTF-ZF)}=   \Hcal_{0,i,i}^\Hrm (\bm{\tau}_{\rm tr}) (  \Hcal_{0,i,i}   \Hcal_{0,i,i}^\Hrm(\bm{\tau}_{\rm tr}) )^{-1}$ in order to slightly better match the desired channel. Nevertheless, $\Wbf^{\rm (wTF-ZF)} \approx \Wbf^{\rm (TF-ZF)}$  for thresholds in the vicinity of the CP, and with equality if all thresholds are equal to the CP.} 
\begin{equation} \label{eq:TF-ZF}
\Wbf^{\rm (TF-ZF)} =   \Hcal_i^\Hrm (\bm{\tau}_{\rm tr}) (  \Hcal_{0,i,i}   \Hcal_i^\Hrm(\bm{\tau}_{\rm tr}) )^{-1}.
\end{equation}

For illustrating the feasibility of TF precoding in a multi-user scenario, the performance of these different ZF precoders~\eqref{eq:F-ZF},~\eqref{eq:F-ZF2} and~\eqref{eq:TF-ZF} are compared in Fig.~\ref{fig:MU} by sum-throughput simulations of  2 and 4 users, for 200 transmit antennas and 64QAM. Each user channel has the same CIR and the truncation threshold is set equal to the CP length for all users. Equal power allocation is assumed and thus the precoders' columns are normalized accordingly. Other simulation parameters are otherwise as in Section VI. In Fig.~\ref{fig:MU}, one verifies that TF-precoding~\eqref{eq:TF-ZF} mitigates more ISI/ICI  and thus provides a higher sum-rate than both the conventional F-precoding~\eqref{eq:F-ZF} and its corrected version~\eqref{eq:F-ZF2}. F-precoding with an extended CP provides as seen before a faster convergence but with a heavily penalized maximum sum-rate due to the CP overhead.  
For illustrating the feasibility of TF precoding in a multi-user scenario, the performance of these different ZF precoders~\eqref{eq:F-ZF},~\eqref{eq:F-ZF2} and~\eqref{eq:TF-ZF} are compared in Fig.~\ref{fig:MU} by sum-throughput simulations of  2 and 4 users, for 200 transmit antennas and 64QAM. Each user channel has the same CIR and the truncation threshold is set equal to the CP length for all users. Equal power allocation is assumed and thus the precoders' columns are normalized accordingly. Other simulation parameters are otherwise as in Section VI. In Fig.~\ref{fig:MU}, one verifies that TF-precoding~\eqref{eq:TF-ZF} mitigates more ISI/ICI  and thus provides a higher sum-rate than both the conventional F-precoding~\eqref{eq:F-ZF} and its corrected version~\eqref{eq:F-ZF2}. F-precoding with an extended CP provides as seen before a faster convergence but with a heavily penalized maximum sum-rate due to the CP overhead.

\section{Conclusions}
We studied the possibility of shortening the effective channel  of a multi-antenna OFDM system with per-subcarrier  precoding, i.e. the  legacy implementation of multi-antenna precoding with OFDM.  
It was shown that introducing time-selectivity in conventional frequency-domain multi-antenna precoders can asymptotically mitigate the self-interference resulting from an insufficient CP as the number of antennas grows large.    
Time-frequency selective precoders were thus constructed based on the truncation of the channel impulse response in order to reduce the  time-dispersion of the effective precoded channel. An optimum threshold in the high-SNR regime was set equal to the CP length in order to achieve an interference-free transmission as the number of antennas goes to infinity. In a medium SNR regime, an optimized threshold not necessarily equal to the CP  further slightly increased the system performance  by accumulating more multi-path energy at the cost of limited interference. 
When a conventional system either suffers from a SINR saturation or a large CP overhead, the performance gain in the high-SNR regime compared to conventional precoders without time-selectivity could be substantial. 
Meanwhile, this design that preserves the legacy frequency-domain precoding structure in OFDM, is asymptotically suboptimal to the time-reversal filtering technique, that can accumulate all path energy and asymptotically convert it to a zero-delay channel, therefore removing the necessity of a CP. It was nevertheless shown that time-frequency selective precoding, which is still exploiting the benefit of a CP to deal with less interference, could converge faster to its asymptotic performance and as a result could even outperform time-reversal filtering for not-so-large array system.

\section*{appendices}

\subsection{Proofs of  Lemma~\ref{lem:F-precoding}}
\label{App:Lemma1}

By direct expansion of $\sbf$ and $\wbf_l$, we get 
\begin{multline}
r[k] = \frac{1}{\sqrt{\Nsc}} \sum_{m=0}^{k+\Ncp}\sum_{l=0}^{\Nsc-1}  \sum_{p=0}^{L-1} \frac{\hbf^\Hrm[p]\hbf[m]}{\omega_l} x_{0,l} e^{\jrm 2 \pi \frac{l(k-m+p)}{\Nfft}}\\ 
+\frac{1}{\sqrt{\Nsc}} \sum_{m=k+\Ncp+1}^{L-1}\sum_{l=0}^{\Nsc-1}  \sum_{p=0}^{L-1}
 \frac{\hbf^{\Hrm}[p]\hbf[m]}{\omega_l} x_{-1,l} \\ \mkern40mu \times e^{\jrm 2 \pi \frac{l(k-m+p+\Nfft+\Ncp)}{\Nfft}}  +z[k]. 
\end{multline}

Since  $\frac{\hbf^{\Hrm}[p]\hbf[m]}{\Nt} \to E_m\delta_{m,p}$ as $\Nt \to \infty$,  we similarly get $\frac{\hbf^{\Hrm}[p]\hbf[m]}{\sqrt{\Nt}\omega_l} \to \frac{E_m}{\alpha_L} \delta_{m,p}$ and 
\begin{eqnarray}
\frac{r[k]}{\sqrt{\Nt}} \mkern-15mu &\to&\mkern-15mu
\frac{1}{\sqrt{\Nsc}} \sum_{l=0}^{\Nsc-1} \sum_{m=0}^{\min(k+\Ncp,L-1)} \frac{E_m}{\alpha_L} x_{0,l} e^{\jrm 2 \pi \frac{lk}{\Nfft}} 
\nonumber \\
& &\mkern-15mu + \frac{1}{\sqrt{\Nsc}}\sum_{l=0}^{\Nsc-1} \sum_{m=k+\Ncp+1}^{L-1} \frac{E_m}{\alpha_L}  x_{-1,l} e^{\jrm 2 \pi \frac{l(k+\Ncp+\Nfft)}{\Nfft}} \nonumber \\
&\mkern-60mu=& \mkern-40mu\frac{\beta_k^2}{\alpha_L}  s_0[k] +\frac{\alpha^2_L-\beta_k^2}{\alpha_L} \frac{1}{\sqrt{\Nsc}}\sum_{l=0}^{\Nsc-1}  x_{-1,l} e^{\jrm 2 \pi \frac{l(k+\Ncp)}{\Nfft}}.
\end{eqnarray}

For the indexes $-\Ncp \leq k<\Nfft-\Ncp$, one has 
$$\displaystyle \frac{1}{\sqrt{\Nsc}}\sum_{l=0}^{\Nsc-1}  x_{-1,l} e^{\jrm 2 \pi \frac{l(k+\Ncp)}{\Nfft}} = s_{-1}[k+\Ncp] ,$$ otherwise if $\Nfft-\Ncp\leq k\leq \Nfft-1$ then $(\alpha^2_L-\beta_k^2) =0$ as $L\leq \Nfft$, so we can write in general for $-\Ncp\leq k\leq \Nfft-1$
\begin{eqnarray}
\frac{r[k]}{\sqrt{\Nt}} &\to& \frac{\beta_k^2}{\alpha_L}  s_0[k] +\frac{\alpha^2_L-\beta_k^2}{\alpha_L} s_{-1}[k+\Ncp] \\
&=& \frac{\beta_k^2}{\alpha_L}  s[k] +\frac{\alpha^2_L-\beta_k^2}{\alpha_L} s[k-\Nfft]. 
\end{eqnarray}
In the case $ L-\Ncp \leq k \leq \Nfft-1$, we have $\beta_k = \alpha_L$.

\subsection{Proof of Proposition 1}
\label{App:Prop1}
From $\frac{\omega_l(\Th)}{\sqrt{\Nt}} \to \alpha_{\Th}$ and then $\frac{\hbf^{\Hrm}[p]\hbf[m]}{\sqrt{\Nt}\omega_l(\Th)} \to \frac{E_m}{\alpha_{\Th}} \delta_{m,p}$,  it is a direct verification that

\begin{eqnarray}
\frac{\hcal^\Trm_{0,i,i} \wbf_i(\Th)}{\sqrt{\Nt}} \mkern-10mu &=& \mkern-10mu\sum_{m=0}^{L-1}\sum_{p=0}^{\Th-1} c[m] \frac{\hbf^\Hrm[p]\hbf[m]}{\sqrt{\Nt} \omega_i(\Th)} e^{-\jrm 2 \pi \frac{i(m-p)}{\Nfft}}\nonumber
\\&\to& 
\frac{1}{\alpha_{\Th}} \sum^{\Th-1}_{p=0} c[p] E_p, 
\end{eqnarray}

\begin{eqnarray}
\frac{\hcal^\Trm_{0,l,i} \wbf_l(\Th)}{\sqrt{\Nt}} \mkern-10mu &=&\mkern-20mu \sum_{m=\Ncp+1}^{L-1}\sum_{p=0}^{\Th-1} \tilde{c}_{l,i}[m] \frac{\hbf^\Hrm[p]\hbf[m]}{\sqrt{\Nt} \omega_l(\Th)} e^{-\jrm 2 \pi \frac{l(m-p)}{\Nfft}} \nonumber
\\&\to& 
 \frac{1}{\alpha_{\Th}} \sum^{\Th-1}_{p=\Ncp+1} \tilde{c}_{l,i}[p] E_p,
\end{eqnarray}
and similarly  
\begin{equation}
\frac{\wbf_i^\Trm(\Th)(\hcal_i-\hcal_{0,i,i}) }{\sqrt{\Nt}} \to \frac{1}{\alpha_{\Th}} \sum^{\Th-1}_{p=\Ncp+1} (1-c[p])  E_p.
\end{equation}
Then the SINR expression follows directly by inserting the above expressions in \eqref{eq:SINRi}.

\vspace{-0.0cm}
\subsection{Derivation Details of the Finite-Size Approximations}

\subsubsection{TF-Precoding}
\label{App:FS_TF}
As discussed in Section V. A., since the numerator and denominator in the SINR  both scaled by $1/\Nt$  are tending to finite deterministic values, we can approximate the SINR by the ratio of the averages as

\begin{equation}
\label{eq:SINRiapprox}
{\rm SINR}_i \approx 
\frac{ \expect{|\hcal^\Trm_{0,i,i} \wbf_i(\Th)|^2}}
{\left( \splitdfrac{\expect{|\left(\hcal^\Trm_i-\hcal^\Trm_{0,i,i}\right)\wbf_i(\Th)|^2}+{\rm SNR}^{-1}}{+ \sum_{\substack{l=0\\l\neq i}}^{\Nsc-1}  2\expect{|\hcal^\Trm_{0,l,i}\wbf_l(\Th)|^2} }\right)}.
\end{equation}

Let us write the precoders without normalization by $\vbf_i(\Th) = \hcal^*_i(\Th)$ such that  $\wbf_i(\Th) = \frac{ \vbf_i(\Th)}{\|  \vbf_i(\Th) \|}$. 
A simple modification of the proof of Prop. 1 shows that $\frac{\hcal^\Trm_{0,i,i} \vbf_i(\Th)}{\Nt}$ and $\frac{\|\vbf_i(\Th)\|}{\sqrt{\Nt}}$ tends to finite and non-zero deterministic values as $\Nt$ becomes large, so we can approximate their ratio by the ratio of their average values as

\begin{multline}
 \frac{\expect{|\hcal^\Trm_{0,i,i} \wbf_i(\Th)|^2}}{\Nt} = 
 \expect{ \frac{|\hcal^\Trm_{0,i,i}\vbf_i(\Th)|^2}{\Nt^2} \times\frac{\Nt}{\|\vbf_i(\Th)\|^2} }
\\ 
\mkern140mu \approx \frac{\expect{|\hcal^\Trm_{0,i,i}\vbf_i(\Th)|^2}}{\Nt^2}  \times\frac{\Nt}{\expect{\|\vbf_i(\Th)\|^2}} 
\\
\approx   
\frac{\expect{|\hcal^\Trm_{0,i,i}\vbf_i(\Th)|^2}}{\Nt \expect{\|\vbf_i(\Th)\|^2}}. \mkern85mu
\end{multline}

We similarly approximate 
{$\displaystyle \frac{\expect{|\left(\hcal^\Trm_i-\hcal^\Trm_{0,i,i}\right)\wbf_l(\Th)|^2} }{\Nt} \approx \frac{\expect{|\left(\hcal^\Trm_i-\hcal^\Trm_{0,i,i}\right)\vbf_l(\Th)|^2} }{\Nt \expect{\|\vbf_l(\Th)\|^2}} $} and 
{$\displaystyle \frac{\expect{|\hcal^\Trm_{0,l,i}\wbf_l(\Th)|^2}}{\Nt} \approx \frac{\expect{|\hcal^\Trm_{0,l,i}\vbf_l(\Th)|^2}}{\Nt \expect{\|\vbf_l(\Th)\|^2}}$}, so that the SINR is further approximated as 

\begin{eqnarray}
 {\rm SINR}_i \mkern-15mu &\approx&\mkern-15mu   
\frac{\displaystyle \frac{\expect{ |\hcal^\Trm_{0,i,i} \vbf_i(\Th)|^2}}{ \expect{\|\vbf_i(\Th)\|^2}}}
{ \left(\splitdfrac{\displaystyle 
\frac{\expect{|\left(\hcal^\Trm_i-\hcal^\Trm_{0,i,i}\right)\vbf_i(\Th)|^2 }}{\expect{\|\vbf_i(\Th)\|^2}} +{\rm SNR}^{-1}
}{\displaystyle 
+ \sum_{\substack{l=0\\l\neq i}}^{\Nsc-1}  2\frac{\expect{|\hcal^\Trm_{0,l,i}\vbf_l(\Th)|^2}}{ \expect{\|\vbf_l(\Th)\|^2}} 
}\right)}  \nonumber
\\&\triangleq& 
  \Gamma_i(\Th) .
\end{eqnarray}

%

Now, only four simple expectations in this expression need to be computed. 
First, it is easy to verify that  $\expect{\|\vbf_l(\Th) \|^2}  = \Nt \alpha_{\Th}^2$ which is independent of $l$. Then, after expansion we obtain

\begin{multline}
\label{eq:EHoii}
\expect{|\hcal^\Trm_{0,i,i} \vbf_i(\Th)|^2} = \sum_{m=0}^{L-1} \sum_{p=0}^{\Th-1} \sum_{m'=0}^{L-1} \sum_{p'=0}^{\Th-1}c[m]c[m'] 
\\\times 
\expect{ \hbf^\Hrm[p]\hbf[m]\hbf^\Hrm[m']\hbf[p']}
e^{-\jrm 2\pi \frac{i(m-p-m'+p')}{\Nfft}}. 
\end{multline}   

For a given $(m,\,p)$, there are only three cases where $\expect{ \hbf^\Hrm[p]\hbf[m]\hbf^{\Hrm}[m']\hbf[p']} \neq 0$: 

\begin{itemize}
\item if  $m=p,\, m'\neq m,\,p' = m'$, then 
\begin{equation}\expect{\|\hbf[m]\|^2 \|\hbf[m']\|^2} = \Nt^2 E_m E_{m'}, \end{equation}

\item if $m=p=m=m'$, then using~\cite[Lem. 1]{LimTWC15} 
\begin{equation} \expect{\|\hbf[m]\|^4}= (\Nt^2+\Nt) E_m^2 ,\end{equation}

\item  if $m\neq p,\, m'= m,\,p' = p$, then 
\begin{equation} \expect{|\hbf^\Hrm[p]\hbf[m]|^2} = \Nt E_m E_p.\end{equation}
\end{itemize}

From which we get 
\begin{align}
&\expect{|\hcal^\Trm_{0,i,i} \vbf_i(\Th)|^2} =\nonumber \\
& \mkern20mu \Nt^2 \sum_{m=0}^{\Th-1}   \sum_{\substack{m'=0\\m'\neq m}}^{\Th-1} c[m] c[m'] E_m E_{m'}  \nonumber\\
& \mkern30mu + (\Nt^2 + \Nt) \sum_{m=0}^{\Th-1} c[m]^2 E_m^2+ \Nt \sum_{m=0}^{L-1}   \sum_{\substack{p=0\\p\neq m}}^{\Th-1} c[m]^2 E_m E_p \nonumber\\ 
&= \Nt^2 \left( \sum_{m=0}^{\Th-1} c[m] E_m\right)^2 \nonumber\\
& \mkern100mu+ \Nt \left(\sum_{p=0}^{\Th-1} c[p]^2 E_p^2 + \sum_{m=0}^{L-1}\sum_{\substack{p=0\\p\neq m}}^{\Th-1} c[m]^2 E_m E_p \right)  \nonumber \\
&= \Nt^2 \left( \sum_{m=0}^{\Th-1} c[m] E_m\right)^2 + \Nt \left(\sum_{p=0}^{\Th-1} E_p \right) \left(\sum_{m=0}^{L-1}  c[m]^2 E_m \right) \nonumber\\
&= \Nt^2 \left( \sum_{m=0}^{\Th-1} c[m] E_m \right)^2 + \Nt \alpha_{\Th}^2 \sum_{m=0}^{L-1} c[m]^2 E_m
\end{align}

Similarly, we obtain 
\begin{multline}
\expect{ |\hcal^\Trm_{0,l,i} \vbf_i(\Th)|^2} = \\
\Nt^2 \left| \sum_{m=\Ncp+1}^{\Th-1} \tilde{c}_{l,i}[m] E_m \right|^2 + \Nt \alpha_{\Th}^2 \mkern-15mu\sum_{m=\Ncp+1}^{L-1}\mkern-15mu | \tilde{c}_{l,i}[m]|^2 E_m,
\end{multline}   
 and 
\begin{multline}
\expect{|\left(\hcal^\Trm_i-\hcal^\Trm_{0,i,i}\right)\vbf_l(\Th)|^2}  = \\\Nt^2 \left( \sum_{m=\Ncp+1}^{\Th-1} (1-c[m]) E_m \right)^2 
\mkern-5mu+ \Nt \alpha_{\Th}^2 \mkern-15mu\sum_{m=\Ncp+1}^{L-1} \mkern-15mu (1-c[m])^2 .
\end{multline}


\subsubsection{TR-filtering}
\label{App:FS_TR}
Since the filtering normalization $\tilde{\omega}$ is the same inside all terms of~\eqref{eq:SINRiTR}, we can rewrite it as
\begin{equation}
\label{eq:SINRiTR2}
{\rm \widetilde{SINR}}_i= 
\frac{ \frac{1}{\Nt^2} \tilde{\Scal}}{\frac{1}{\Nt^2} \tilde{\Ical} + \frac{|\tilde{\omega}|^2}{\Nt} {\rm SNR_{op}^{-1}}}
\end{equation}
with $\tilde{\Scal} =  |\tilde{\gcal}_{0,i,i}|^2 $ and $\tilde{\Ical} = \sum_{\substack{l=0\\l\neq i}}^{\Nsc-1}  |\tilde{\gcal}_{0,l,i}|^2 +\sum_{l=0}^{\Nsc-1}  \left(|\tilde{\gcal}_{-1,l,i}|^2 +|\tilde{\gcal}_{1,l,i}|^2\right) $ 
where $\tilde{\gcal}_{b,l,i}= \tilde{\omega} \gcal_{b,l,i}$ are equivalently expressible as in~\eqref{eq:G0ii}--\eqref{eq:Gbli} but for the unnormalized channel $\tilde{g}[n]=  \sum_{m=0}^L \hbf^{\Hrm}[m-n]\hbf[m]$ instead of $g[n]$. 
It can be verified that each terms in~\eqref{eq:SINRiTR2} tends to finite values as $\Nt \to \infty$ and so 
\begin{equation}
\label{eq:SINRiTR3}
{\rm \widetilde{SINR}}_i \approx \frac{\expectl{\tilde{\Scal}}}{ \expectl{\tilde{\Ical}} + \Nt\expect{|\tilde{\omega}|^2} {\rm SNR_{op}^{-1}}}.
\end{equation}

First we have $\expect{|\tilde{\omega}|^2} = \Nt \alpha_L^2$. 
Then, defining the average channel power $\tilde{E}_m = \expect{\tilde{g}[m]}$, and using similar identities than before,
we find 
\begin{equation}
\tilde{E}_m  = \Nt  \left(  \sum_{n=0}^{L-1}  E_n E_{n-m}  +  \Nt \alpha_L^4 \delta_{0,m} \right). 
\end{equation}

From~\eqref{eq:G0ii}--\eqref{eq:Gbli}, we have
\begin{equation}
\expectl{\tilde{\Scal}} = \sum_{m=-L+1}^{L-1} c[m+\Delta]^2 \tilde{E}_m 
 \end{equation}
and 
\begin{align}
&\expectl{\tilde{\Ical}}=  \nonumber \\
&\quad \sum_{m=-L+1}^{L-1} \left( (1-c[m+\Delta])^2 
+  \sum_{\substack{l=0\\l\neq i}}^{\Nsc-1} 2 |\tilde{c}_{l,i}[m+\Delta]|^2   \right) \tilde{E}_m  \nonumber\\
&= \sum_{m=-L+1}^{L-1} \left( \sum_{l=0}^{\Nsc-1} 2 |\tilde{c}_{l,i}[m+\Delta]|^2   \right) \tilde{E}_m. \label{eq:E[I]}
\end{align}

It follows that the SINR is approximated as
\begin{multline}
\label{eq:SINRiTR4}
{\rm \widetilde{SINR}}_i \approx \\
\frac{\sum_{m=-L+1}^{L-1} c[m+\Delta]^2 \tilde{E}_m  }
{ \sum_{m=-L+1}^{L-1} \left( \sum_{l=0}^{\Nsc-1} 2 |\tilde{c}_{l,i}[m+\Delta]|^2   \right) \tilde{E}_m + \Nt^2 \alpha_L^2  {\rm SNR_{op}^{-1}}}
\end{multline}
which simplifies to~\eqref{eq:SINRiTRapprox} by defining $\rho_m =  \sum_{n=0}^{L-1}  E_n E_{n-m}$ and consecutively 
$\tilde{E}_m  = \Nt  \rho_m   +  \Nt^2 \alpha_L^4 \delta_m  $.

\emph{Special Case $\Nsc = \Nfft$:} the following identity can be verified by expansion~\cite{PitavalArxiv21}
\begin{equation}
\sum_{\substack{l=0\\l\neq i}}^{\Nsc-1} 2 |\tilde{c}_{l,i}[m]|^2  = c[m]-c[m]^2, 
 \end{equation}
and in this case we have the simplification of~\eqref{eq:E[I]}
to 
\begin{equation}
\expectl{\tilde{\Ical} } =  \sum_{m=-L+1}^{L-1}   (1-c[m+\Delta]^2)\tilde{E}_m . 
 \end{equation}

\addcontentsline{toc}{chapter}{Bibliography}
\bibliographystyle{IEEEtran} 
\bibliography{mybibfile}

\end{document}